\documentclass[%
 reprint,
 amsmath,amssymb,
 aps,
]{revtex4-2}
\UseRawInputEncoding

\usepackage[british]{babel}
\usepackage{graphicx}
\usepackage{dcolumn}
\usepackage{bm}
\usepackage[colorlinks,linkcolor=blue,citecolor=blue,urlcolor=blue]{hyperref}
\usepackage{xcolor}
\usepackage[normalem]{ulem}
\def\comment#1{}

\def\Re{{\rm Re\,}}

\def\dbar{{\mathchar'26\mkern-12mu d}} 
\def\ttheta{\theta}
\def\theta{\vartheta}
\def\omega{w}

\usepackage{tikz}
\definecolor{darkgreen}{rgb}{0,0.6,0.2}

\newtheorem{theorem}{Theorem}

\newtheorem{definition}[theorem]{Definition}

\newcommand{\beq}{\begin{equation}}
\newcommand{\eeq}{\end{equation}}
\newcommand{\bea}{\begin{eqnarray}}
\newcommand{\eea}{\end{eqnarray}}

\definecolor{darkred}{rgb}{.8,0,0}

\definecolor{darkblue}{rgb}{0,0,0}

\definecolor{bluec}{rgb}{0,0,0.7}

\begin{document}

\preprint{APS/123-QED}

\title{Universality classes, Thermodynamics of Group Entropies, and Black Holes}

\author{Henrik Jeldtoft Jensen}
 \email{h.jensen@imperial.ac.uk}
\affiliation{
Centre for Complexity Science and Department of Mathematics, Imperial College London, South Kensington Campus, London SW7 2AZ, UK 
}%

\author{{Petr Jizba}}
 \email{p.jizba@fjfi.cvut.cz}
\affiliation{Faculty of Nuclear Sciences and Physical Engineering, Czech Technical University in Prague, B\v{r}ehov\'{a} 7, 115 19, Prague, Czech Republic
}

\author{{Piergiulio Tempesta}}
 \email{p.tempesta@fis.ucm.es}
\affiliation{Departamento de F\'{\i}sica Te\'{o}rica II (M\'{e}todos Matem\'{a}ticos de la F\'isica), Facultad de F\'{\i}sicas, Universidad
Complutense de Madrid, 28040 -- Madrid, Spain \\ and
Instituto de Ciencias Matem\'aticas, C. Nicol\'as Cabrera, No 13--15, 28049 Madrid, Spain
 \vspace{0.5cm}
}

\date{6 October 2026}

\begin{abstract}
Conventional Boltzmann--Gibbs statistical mechanics successfully describes systems with weak to moderate correlations, in which the number of accessible configurations $W(N)$  grows exponentially with the number of degrees of freedom~$N$. However, this framework breaks down for systems for which the effectively occupied configuration space  exhibits non-exponential growth. While numerous generalized entropies have been proposed to address this limitation, a coherent link to classical thermodynamic laws has remained elusive. Here, we propose group entropies as a unifying framework, defining universality classes of entropies through the asymptotic scaling of 
$W(N)$, each yielding an extensive entropy. 
We show that this approach provides the basis for a consistent thermodynamic formulation beyond the Boltzmann--Gibbs paradigm.
In particular, by expressing these entropies in terms of thermodynamic state variables, we demonstrate their consistency with classical thermodynamics, in close analogy to the emergence of the Clausius entropy from the Boltzmann--Gibbs formalism. Focusing on the zeroth thermodynamic law, we identify the empirical temperature and, by using Carath\'{e}odory's formulation of the second law, we derive the associated absolute temperature. As an application, we analyse black-hole thermodynamics using the extensive group entropy class corresponding to stretched-exponential behaviour of $W(N)$ and derive a negative heat capacity, known to be a hallmark property of black holes. This result holds for the stretched-exponential entropies associated with both the Bekenstein--Hawking and Barrow entropy scalings.

\end{abstract}

\maketitle


\section{Introduction \label{Intro}}

Classical thermodynamics is a phenomenological theory of macroscopic systems, grounded in experimental observations and summarized in the laws of thermodynamics (see, e.g., Ref.~\cite{Huang}). It describes heat transformations and the resulting changes in a system in terms of macroscopic state variables, such as energy, volume, temperature, and, for example, magnetic field, without reference to the underlying microscopic constituents. In this framework, entropy, as introduced by Clausius, is treated as a function of the macroscopic state variables.

Statistical mechanics, as originally developed by Boltzmann, Gibbs, and others, aspires to derive thermodynamics from statistical arguments based on the average or most probable (or most typical) behaviour of the microscopic constituents. In this context, the Boltzmann--Gibbs entropy is expressed as a functional on the space of probability distributions over the possible microscopic configurations:  
\begin{equation}
	S_{\rm BG}(P) \ = \  k_{\rm B} \sum_{i=1}^{W(N)} p_i \ln \frac{1}{p_i}\, .
	\label{BG_ent}
\end{equation}
Here, $k_{\rm B}$ denotes Boltzmann's constant (for ease of notation we use $k_{\rm B}=1$ from now on), $N$ denotes the number of constituent parts --- particles or degrees of freedom --- in the system, and $W(N)$ denotes the total number of microscopic configurations. Each microscopic configuration, labeled by $i$, occurs with probability $p_i$. 
In what follows, we denote by $\mathcal{P}$ the set of all probability distributions $P = (p_1, \dots, p_W)$. Assuming that the observed macroscopic behaviour corresponds to the most likely configurations, one can establish mathematical statements equivalent to the zeroth law of thermodynamics (equilibrium is reached when two systems in thermal contact share the same temperature) and the first law (energy is conserved and expressed through exchanged work and heat) by analysing the entropy in Eq~\eqref{BG_ent} in the microcanonical ensemble, where all microstates contribute with the same probability $p_i=1/W(N)$, see e.g. Refs.~\cite{Huang,Reif1965}.


Boltzmann first demonstrated through his $H$-theorem  that the expression for the entropy given by~(\ref{BG_ent}) is consistent, for a dilute gas, with the second law of thermodynamics (i.e., entropy does not decrease in an isolated system), provided the hypothesis of molecular chaos (Stosszahlansatz) is imposed~\cite{Boltzmann}. This assumption amounts to neglecting correlations at the level of binary collisions and therefore constitutes a statistical coarse-graining. However, it was Gibbs who rigorously formulated Eq.~(\ref{BG_ent}) as an entropy defined on phase space and, through the introduction of ensemble theory, established its connection with Clausius' thermodynamic entropy~\cite{Gibbs}. Gibbs also recognized that for an isolated Hamiltonian system the fine-grained entropy remains constant due to Liouville's theorem, and so any effective realization of the second law requires a coarse-grained description.

The entropic form~(\ref{BG_ent}) was later rediscovered by Shannon in the context of information theory as a quantitative measure of uncertainty~\cite{Shannon}. Jaynes subsequently pioneered the use of Shannon entropy in statistical physics~\cite{Jaynes} by promoting the {Maximum Entropy principle} (MaxEnt) to a general variational framework for deriving equilibrium ensembles. The transition from Shannon's information theory to equilibrium statistical mechanics is formally straightforward: maximizing entropy while considering constraints on average energy, or energy and particle number,  yields canonical or grand canonical Gibbs distributions, respectively. In this sense, MaxEnt --- already implicitly used by Gibbs in his principle of the most probable distribution --- provides the bridge that elevates entropy from a functional on probability space to a thermodynamic state function.

The impressive success of the Boltzmann--Gibbs statistical mechanics is, however, limited to situations where the number of allowed configurations grows exponentially with the number of constituents, i.e., when $W(N)\propto \exp(N)$. In this case, the Boltzmann--Gibbs entropy is both additive and extensive. Additivity means that for two statistically independent systems $A$ and $B$, the entropy of the combined system $A\times B$, formed as the Cartesian product of the set of states of $A$ and $B$, satisfies $S(A\times B)=S(A)+S(B)$. Extensivity can be expressed as $S(N)\sim N$ asymptotically when $N$ goes to infinity.
For systems where $W(N)$ deviates from exponential growth --- for instance,  $W(N)\sim N^a$ or $W(N)\sim \exp(N^\gamma)$ --- the Boltzmann--Gibbs entropy  fails to be extensive, and its standard thermodynamic meaning breaks down.


For completeness, we mention a few examples in which the 
asymptotic growth of $W(N)$ deviates from the standard exponential form. The pairing model discussed in Ref.~\cite{JPPT2018} consists of particles that can occupy internal states and additionally form dynamically generated paired states, leading to a super-exponential growth law of the form $W(N)\sim N^{a N}$. Ref.~\cite{HanelTHurner2011} classifies a number of stochastic processes whose associated state spaces exhibit non-exponential scaling laws for $W(N)$, including accelerating random walks characterized by algebraic growth $W(N)\sim N^{a}$. 

A particularly important and well-known example of sub-exponential but non-trivial growth arises in hadronic bootstrap models and string theory~\cite{Frautschi1971,GreenSchwarzWitten1987,Polchinski1998}, where Hagedorn-type spectra lead asymptotically to $W(N)\sim \exp(a\sqrt{N})$ ($a>0$ and $N$ is the total oscillator number). This behaviour follows from the combinatorics of constrained collective excitations, and provides a canonical microscopic realization of how global constraints can strongly modify state-space growth.

Finally, in the context of ordinal analysis of data sequences, super-exponential growth typically arises in the study of signals generated by stochastic sources, where the relevant state space corresponds to permutations of temporal orderings~\cite{BandtPompe2002}.

It should be stressed that extensivity is essential to ensure that the entropy per particle (or effective degree of freedom) exists, or in other words, that the limit
\begin{eqnarray}
\lim_{N\rightarrow \infty} \frac{S_{N}(E_N)}{N}\, ,
\end{eqnarray}
is finite. 
Hence, the entropic extensivity is the target property one should attempt to enforce or recover.
By contrast, the finiteness of the energy per particle is not guaranteed in general. In particular, for systems with long-range interactions, this quantity may diverge; a text-book example is a single electron together with its
own electrostatic field, whose total energy is infinite,  or a self-gravitating gas, whose energy at fixed density grows as $N^{5/3}$ \cite{Rufo:14}.

Entropic additivity is less essential, but we still need a systematic rule for combining statistically independent systems. More specifically, we need the entropy to be composable, which guarantees that we obtain the same expression for the combined system irrespective of whether we compute the entropy of $A\times B$ as one system, or we first compute the entropy of $A$ and of $B$ separately and then combine them. This implies that any generalization of the Boltzmann--Gibbs entropic functional form in Eq.~(\ref{BG_ent}) must be both composable and extensive. 
A wealth of generalized entropic functionals have been proposed in the literature (see, e.g. Ref.~\cite{Hanel} and citations therein). Notable examples include the Tsallis entropy~\cite{Tsac}, which is appropriate for systems whose number of accessible microstates scales as $W(N)\sim N^{a}$, and the entropy introduced in Ref.~\cite{Tempesta} for systems exhibiting stretched-exponential growth, $W(N)\sim \exp(N^{\gamma})$. 
The latter has been suggested to be relevant in the cosmological context~\cite{T-J}. These two entropies are both composable and extensive for their respective $W(N)$ scaling behaviours, whereas other suggested entropies are not. For instance, the popular $\delta$-entropies~\cite{Tsac,Tsallis:2012js,Tsab,Tsallis_epl_2026, BeckTsallis2026} given by 
\begin{eqnarray}
S_\delta(P) \ = \ \sum_{i=1}^W p_i \left( \ln \frac{1}{p_i}\right)^{\!\!\delta}\, , \;\;\;\; \delta \ > \ 0\, ,
\label{I.1.cc}
\end{eqnarray}
and their generalizations have recently found applications in diverse areas ranging from cosmology to turbulence. However, from an axiomatic standpoint, they are not composable. While their phenomenological and empirical value is well-established, this lack of composability precludes their direct use within the framework of statistical thermodynamics formulated here. 


The framework of group entropies, introduced by one of us in Refs.~\cite{PT2011PRE,T2016AOP,Tempesta2016}, provides a systematic axiomatic generalization of the Shannon--Khinchin framework underlying the Boltzmann--Gibbs entropy. It also furnishes a general constructive procedure for defining composable and extensive entropic functionals corresponding to a given asymptotic scaling of the number of accessible microstates $W(N)$.
This is done by making use of formal group theory, which ensures the existence of a consistent composition law expressed in terms of a group generator $G(t)$.
The latter is determined in terms of $W(N)$ through the requirement that the resulting group entropy is extensive. Furthermore, since the group entropies derived in this way satisfy the Shore--Johnson axioms of inference theory~\cite{SJ1,SJ2,J-K:19}, they can be used as information-theoretic tools for statistical inference within the MaxEnt framework. This feature is of particular importance, as the MaxEnt prescription typically provides  a bridge between entropic functionals and full-fledged thermodynamic entropies, which act as state functions for thermodynamic systems.    

It is of interest to note that the group entropies maximized under appropriate constraints lead to the often observed $q$-exponential distributions for all classes of $W(N)$ asymptotic behaviour. Hence, observation of $q$-exponential distributions is not limited to the algebraic $N$ dependence corresponding  to the Tsallis entropy, see Sec. 4 in \cite{J-Temp}.

The connection between group entropies and the thermodynamic (Clausius-type) entropy is not \emph{a priori} evident, as it is unclear whether the heat one-form admits an exact differential completion in terms of the group entropy. Even if such a completion exists, it remains uncertain whether the physical temperature can be identified with the integrating factor of the heat one-form. In what follows, we address these issues in two complementary ways. 

First, we consider the statistical derivation of thermodynamic laws. We analyse the statistical-mechanical framework induced by the group entropic functional on probability space, restricted to the uniform (microcanonical) probability distribution $p_i = 1/W(N)$. This is a straightforward generalization of the standard textbook discussions within the Boltzmann--Gibbs framework of statistical thermodynamics, see e.g. Sec.~3.9 in Ref.~\cite{Reif1965}. In this approach one obtains thermodynamic relations by maximizing the number of corresponding microstates. To clear up the meaning of the group entropy temperature we turn to macroscopic thermodynamics. 

So secondly we focus on the thermodynamic limit $N \to \infty$ and analyse the group entropy as a state function defined on macroscopic variables such as energy and pressure. Consistency with the laws of thermodynamics is demonstrated within the framework of Carath{\'e}odory's axiomatic formulation~\cite{Caratheodory}. Interestingly, while the first law --- relating entropy to heat exchange and work --- retains its standard form, the zeroth law is modified: the empirical temperature that characterizes equilibrium between different (sub-)systems is no longer given by the derivative of the group entropy with respect to energy.

After discussing the general thermodynamic behaviour of group entropies, we narrow our analysis to a subclass of special interest for black-hole thermodynamics  --- the stretched-exponential case, defined by
\begin{eqnarray}
W(N) \ \sim \  \exp(N^\gamma)\, , 
\label{3.kkl}
\end{eqnarray}
and the associated form of the group entropy
\begin{eqnarray}\label{SE}
S_{\alpha, \gamma}(A) \ &\equiv& \ S_{\alpha, \gamma}(P^{(A)}) \nonumber \\[2mm] &=& \  \left[ \frac{\ln \sum_{i=1}^{W} \big(p_i^{(A)}\big)^{\alpha}}{1-\alpha}\right]^{\!1/\gamma} \, .
\label{3.kk}
\end{eqnarray}
This group entropy was introduced in Ref.~\cite{Tempesta} and subsequently employed in Ref.~\cite{T-J} for the analysis of cosmological data. It is also directly relevant to black-hole physics. In fact, for $\alpha=\gamma=1$, $S_{\alpha,\gamma}$ reduces to the Boltzmann--Gibbs entropy and, under holographic state counting, reproduces in the microcanonical framework the Bekenstein--Hawking area-law entropy. Moreover, $S_{\alpha,\gamma}$ is closely connected to the Barrow (or Barrow--Tsallis) entropy~\cite{Barrow,Tsallis:2012js}:
\begin{equation}
S_{{\mathrm{B}}} \ \propto \  \left(\frac{A}{4l^2_{\rm P}}\right)^{1+\frac{\Delta}{2}} \, ,
\label{barrow}
\end{equation}
where $l_{\rm P}$ is the Planck length and $\Delta$ ($0\leq \Delta \leq 1$) parametrizes an effective fractal deformation of the horizon expected from quantum-gravity effects. Within a quantum field-theoretic description, $\Delta$ can be interpreted as an anomalous dimension~\cite{JL,JLLM:24}.
In this latter case, the Hausdorff fractal dimension of the horizon becomes $D_{\mathrm{H}} = 2 + \Delta$.

Assuming holographic state counting, the entropy~\eqref{SE}, upon identifying
\begin{equation}
\frac{1}{\gamma} \ = \  1 \ + \ \frac{\Delta}{2}\, ,
\end{equation}
can be regarded as the most general thermodynamically consistent entropic extension that provides a statistical realization of the Barrow entropy~(\ref{barrow}) in the microcanonical regime.  This perspective enables a systematic treatment of several aspects of black-hole thermodynamics from a rather unconventional yet illuminating point of view. Alternatively, and more speculatively (see, e.g., Ref.~\cite{Majumdar_2025}), one may attempt to fix the parameter $\gamma$ by assumptions about the distribution of effective microstates within the black-hole interior, which in that scenario suggests $\gamma=2/3$ (see also Sec.~\ref{VII.cv}). In this sense, the entropy~\eqref{SE} provides a flexible framework: being extensive on the stretched-exponential class~\eqref{3.kkl}, it allows one to extract thermodynamic properties of black holes (such as the heat capacity and Stefan--Boltzmann law) across different physical regimes. 

Beyond the specific context of black-hole thermodynamics, the aim of this work is to show how the corresponding thermodynamic relations follow from general principles that are applicable to the full class of group entropies.

The structure of the paper is as follows. In Section~\ref{Sec.II}, we outline the general framework of group entropies, providing the basic definitions, relations, and transformation rules used in the bulk of the paper. 

Section~\ref{Section VI} is devoted to deriving the equilibrium condition and the first thermodynamic law from statistical considerations of the microstates within the microcanonical ensemble for a general group entropy. The analysis indicates a need for a deeper thermodynamical analysis of temperature within the group entropic formalism.

This is done in Section~\ref{Sec.III.aaa}, where we turn to macroscopic thermodynamics and consider the consistency of group entropies as thermodynamic state functions ignoring the micro level. We examine the zeroth law of thermodynamics. In particular, both the empirical temperature and the physical pressure are derived using the concepts of thermal and mechanical equilibrium, respectively.  In Section~\ref{second-law}, we discuss Carath\'{e}odory's formulation of the second law of thermodynamics for systems described by group entropies. We demonstrate that the ensuing integrability conditions indeed allow for the existence of an integrating factor for the heat one-form. This integrating factor factorizes into two components: one depending solely on the empirical temperature and another on the entropy. However, unlike in conventional additive thermodynamics, the entropic component of the integrating factor cannot, in general, be set to a constant and must be accounted for. In Section~\ref{IV.cf}, we further show that the empirical-temperature-dependent component of the integrating factor can be identified with the absolute temperature, as it is independent of the thermometer's material and the nature of the heat bath, and is consistent with the zeroth law of thermodynamics. 

In Section~\ref{VII.cv}, we show that the thermodynamics of the stretched-exponential entropy formally reproduces black hole thermodynamics with a negative heat capacity in the microcanonical ensemble. We further demonstrate that the resulting black-body radiation law implies an absolute temperature that depends on the entropic parameters, leading to a two-parameter family of generalized Stefan--Boltzmann laws.

Finally, in Section~\ref{conclusions}, we summarize the main results obtained and outline possible directions for future research. Additional technical details concerning group entropies and formal proofs of several statements are given in the accompanying Appendices.

\section{Group entropies \label{Sec.II}}
%
For the sake of consistency, we will now briefly review the key aspects of group entropies and explain how composability and extensivity are ensured. This will be needed in the subsequent sections. A more detailed exposition can be found, for example, in Refs.~\cite{PT2011PRE,Tempesta2016,T2016AOP}. 

The composability principle states that the entropy of a composite system is entirely determined by the entropies of its constituent subsystems through a universal composition rule,  $\Phi$. This ensures that entropy does not depend on hidden microscopic details of how the systems are put together, but only on their macroscopic entropies.  In mathematical terms, composability asserts the existence of a symmetric, monotonically increasing function $\Phi(x,y)$ such that for two independent systems $A$ and $B$ (cf. Refs.~\cite{Tsac,T-J,J-Temp,Tempesta2016,Tempesta})
\begin{eqnarray}
S(A\times B) \ = \ \Phi(S(A),S(B))\, .
\label{I.5.cc}
\end{eqnarray}
We emphasize that here $A\times B$ denotes the system obtained as a Cartesian product of all states available to $A$ with those available to $B$. This means that states representing physical interactions between states in $A$ and states in $B$ are not included in $A\times B$. To ensure that the entropy possesses the fundamental structural features demanded by thermodynamic and information-theoretic frameworks, we require compliance with the first three Shannon--Khinchin axioms~\cite{Kchinchin}: (SK1) continuity, (SK2) maximality for the uniform distribution, and (SK3) invariance upon including events of zero probability. Furthermore, for entropy to consistently represent a state function, the composition rule $\Phi(x,y)$ must meet further criteria~\cite{Johal,Scarfone,T2016AOP,Tempesta2016,Hotta,PT2011PRE}. In particular, in~\cite{T2016AOP}, these criteria were formalized as the so-called composability axiom, which generalizes the fourth Shannon--Khinchin axiom, i.e., the additivity axiom.
\begin{definition}
\label{composab} \label{Def1} We say that a generalized entropy $S$ is \textit{(strongly)
composable} if there exists a continuous function of two real
variables $\Phi (x,y)$, called the composition law, such that the following
four conditions are satisfied.

\begin{description}
\item[(C1)] Composability: 
\beq
S(P_1\times P_2)\ = \ \Phi (S(P_1),S(P_2)) \hspace{3mm} \text{\rm {for any}} \hspace{3mm} P_1,P_2\in 
\mathcal{P}\, . \label{con:com}
\eeq
\item[(C2)] Symmetry: 
\beq 
\Phi (x,y) \ = \ \Phi (y,x)\, . \label{con:symm}
\eeq
\item[(C3)] Associativity: 
\beq
\Phi (x,\Phi (y,z)) \ = \ \Phi (\Phi (x,y),z)\, . \label{I.6.cv}
\eeq

\item[(C4)] Null-composability: 
\beq
\Phi (x,0)\ = \ x\, . \label{I.7.cf}
\eeq
\end{description}
\end{definition}
Observe that the mere existence of a function $\Phi
(x,y)$ taking care of the composition process as in condition (C1) is
necessary, but not sufficient to ensure that a given entropy may be suitable
for information-theoretical or thermodynamic purposes; this function must also satisfy the other requirements above to be admissible.
First, it must be \emph{associative}
%
%
so that the entropy of a multi-part system is independent of the order in which subsystems are combined. 
Second, it must be \textit{commutative} 
to guarantee that the compound system does not depend on the ordering of the two subsystems.
Third, it must satisfy \emph{null-composability}
%
%
ensuring that the entropy remains unchanged when a subsystem with zero entropy is added.
A fundamental problem is to characterize all functions $\Phi(x,y)$ which satisfy these nontrivial requirements. The mathematical answer is provided by formal group theory \cite{Haze}, a branch of modern algebraic topology which has found many applications, ranging from number theory to combinatorics, and recently, statistical mechanics. In summary, the three requirements above are equivalent to saying that $\Phi(x,y)$ must be a \textit{commutative formal group law} (see Appendix A). 
This key observation naturally leads to the definition of a group entropy.
\vspace{.3cm}
\begin{definition}
\label{def:groupentropy} A group entropy is a function $S:\mathcal{P}%
\rightarrow \mathbb{R}^{+}\cup \{0\}$ which satisfies the Shannon--Khinchin
axioms (SK1)-(SK3) and the composability axiom (C1)-(C4).
\end{definition}
The substitution of additivity by composability in the four Shannon--Khinchin axioms allows us to generalize the concept of entropy in a way coherent with thermodynamics, as we will show.
Moreover, it can be shown that the Z-entropies satisfy the Shore--Johnson axioms of inference theory~\cite{SJ1,SJ2,J-K:19}.

A crucial mathematical result of formal group theory is that for any formal group law $\Phi(x,y)$ there exists a formal power series $G(t)$ such that
\beq \label{eq:G}
\Phi(x,y) \ = \ G\big(G^{-1}(x) \ + \ G^{-1}(y)\big)\, .
\eeq


In this paper, $G(t)$ will be a continuous, invertible map of $\mathbb{R}_{\geq 0}$ onto itself, not necessarily analytic at the origin, with $G(0)=0$; thus, $\Phi(x,y)$ will be strictly increasing with respect to both arguments.
Generalized logarithms are a key ingredient in the construction of group
entropies and allow us to realize the composition laws for the entropies in
terms of formal group laws. An important instance of the family of group
entropies is the class of $Z$-entropies defined in~\cite{Tempesta2016}. Their
general form, for $\alpha >0$, is 
\begin{equation}
S_{\mathcal{G},\alpha }(P)\ \equiv\ \ \!\frac{1}{1-\alpha }\log _{\mathcal{G}}\Big(\sum_{i=1}^{W}p_{i}^{%
\alpha }\Big)\, ,  \label{eq:Z}
\end{equation}
where $\log _{\mathcal{G}}$ is a generalized logarithm, which for the purposes of this paper can be written in the form $\log _{\mathcal{G}}(x)\equiv \mathcal{G}(\ln x)$ . For the Z-entropies~\eqref{eq:Z} the generator of the formal group law in Eq.~\eqref{eq:G} is related with the function $\mathcal{G}$ entering $\log_{\mathcal{G}}$ via the rescaling $G(t)=\frac{\mathcal{G}((1-\alpha)t)}{1-\alpha}$.


Note that all group entropies automatically satisfy Falk's formulation of the third law, namely that the entropy assumes its minimum value in the state of minimum energy, provided that the ground state is non-degenerate. Indeed, in the ground state only a single microstate is occupied, so the corresponding probability distribution is degenerate, i.e., $p_1=1$ and $p_i=0$ for all $i\neq 1$. In this case, the entropy vanishes because the function $\mathcal{G}(t)$ satisfies $\mathcal{G}(0)=0$. Moreover, since $\mathcal{G}(t)$ is increasing, this value is the global minimum of the entropy.

The extensivity requirement determines the functional form of $\mathcal{G}(t)$ by demanding that the entropy $S_{\mathcal{G},\alpha }(p)$ evaluated on the uniform distribution $p_i=1/W(N)$  satisfies
\begin{equation}
    S_{\mathcal{G},\alpha }({1}/{W(N)}) \ \sim \  \lambda N \;\;\;{\rm for}\;\;\;  N\rightarrow\infty\, ,
    \label{ext_cond}
\end{equation}
where $\lambda$ is a constant independent of $N$.
Once $W(N)$ is given, the group theoretical structure associated with the entropy is known, and vice versa. Namely, the following formula holds:
\begin{equation}
\mathcal{G}(t) \ = \ \lambda (1-\alpha )\left[ W^{-1}\left( e^{\frac{t}{1-\alpha }%
}\right) \ -  \ W^{-1}(1)\right] .  \label{eq:Gt}
\end{equation}
Here, $W^{-1}$ denotes the inverse function of $W$, which we assume to be extended to a continuous, strictly increasing function of a real variable $N\geq 0$.
Thus, the group law associated with $S_{\mathcal{G},\alpha }(p)$ is given by formula \eqref{eq:G} or, equivalently, by
\begin{widetext}
\begin{equation}
\Phi (x,y) \ = \ \lambda \left\{ W^{-1}\Big[W\left( \frac{x}{%
\lambda }+W^{-1}(1)\right) W\left( \frac{y}{\lambda }+%
W^{-1}(1)\right) \Big] \ - \ W^{-1}(1)\right\} \ .
\label{phi_W}
\end{equation}
\end{widetext}
%
%
%



{
Our particular focus here will be on the group entropy based on the stretched-exponential form of  $W(N)$ given by Eq.~(\ref{3.kkl}). Sub-exponential state-space growth of this kind, i.e., the case $\gamma<1$, is observed in systems with long-range interactions or strong  correlations, extending across both classical and quantum domains. Examples of such non-Boltzmann--Gibbs statistics include strongly entangled quantum many-body systems~\cite{Ding:2008,Li:2006}, frustrated spin systems and spin glasses at low temperatures~\cite{KLich:2014,Andrews:2009,Fisch:2006}, as well as cosmological and black-hole systems~\cite{Bekenstein,Hawking,Hawking2}.  Furthermore, such scaling behaviour also characterizes black holes in modified gravity theories, where higher-curvature or higher-dimensional terms are considered~\cite{FJL:26,Estrada:2020,Kubiznak:2023}.

In the \emph{microcanonical ensemble}, where all states are equally probable ($p_i = 1/W$), the corresponding group entropy~(\ref{3.kk}) reduces to
\begin{eqnarray}
S_{\alpha, \gamma} \;=\;  \, (\ln W)^{1/\gamma} \,,
\label{I.10.kl}  
\end{eqnarray}
independent of $\alpha$. Here, $\gamma$ is determined by the scaling relation $W(N) \sim \exp(N^\gamma)$. In the special case $\gamma = 1$, the stretched-exponential entropy~(\ref{3.kk}) reduces to the R\'enyi entropy~\cite{Renyi}, i.e., the non-trace form of group entropy associated with exponential growth of $W(N)$, which is known to be composable. Moreover, in the limit $\gamma\rightarrow 1$, the standard Boltzmann--Gibbs entropy is recovered from Eq.~(\ref{3.kk}) when $\alpha \rightarrow 1$.}

%
%
%

It is worth noting that the entropies $S_{\alpha,\gamma}$ belong to the Uffink class of entropic functionals~\cite{Uffink:95}. Consequently, they can be used consistently in statistical inference procedures implemented within the MaxEnt framework. In the following sections, we further show that the entropies $S_{\alpha,\gamma}$ can be consistently regarded as thermodynamic \emph{state functions}. Accordingly, they provide a potentially useful basis for the thermodynamic description of systems exhibiting stretched-exponential scaling of the state space.

While composability and extensivity serve as the fundamental prerequisites for the generalized macroscopic thermodynamics developed in this work, other physical and information-theoretic applications may prioritize different mathematical criteria. For example, if entropy is interpreted strictly as the standard linear statistical average (the arithmetic expectation value) of a ``surprise'' function associated with individual events, the entropy functional is required to possess a trace form,
$S=\sum_i f(p_i)$.  Notably, the R\'enyi entropy can still be interpreted as a meaningful average of individual surprises, $I(p_i)=-\ln p_i$, provided the standard linear expectation value is replaced by a quasi-linear Kolmogorov--Nagumo mean. In this context, we mention that it has been proven in~\cite{EncisoTempesta2017}  that, under the corresponding axiomatic requirements, the Tsallis entropy emerges as the unique admissible entropic functional that is both strictly composable and of trace form.


\section{Thermodynamics from the microcanonical ensemble of the group entropies } \label{Section VI}

\subsection{Deriving the equilibrium condition using the micro-canonical group entropy}
%

Recall that $W(N)$ denotes the functional form of dependence of number of accessible microstates on $N$. Using Eqs.~(\ref{eq:Z}) and~(\ref{eq:Gt}), the group entropy evaluated in the microcanonical ensemble takes the form $(\lambda=1)$
:
\begin{equation}
	S^{\rm uni}_\alpha \ = \ \big[W^{-1}(\Omega(E)) \ - \ W^{-1}(1)\big]\, .
	\label{S_uni}
\end{equation}
Here, $\Omega(E)$ denotes the actual number of available microstates consistent with the energy $E$. 
As in the standard microcanonical treatment (see, e.g., Ref.~\cite{Reif1965}, Sec.~3.3), we may assume that systems $A$ and $B$ are placed in thermal contact while the composite system remains thermally isolated. The interaction between $A$ and $B$ is assumed to be so weak that the set of states for the combined system $A\times B$ is given by the Cartesian combination of the microstates of the two systems. That is, the total number of states of the combined system is
\begin{equation}
	\Omega^{\rm Tot}(E^{\rm Tot}) \ = \ \Omega_A(E_A)\Omega_B(E^{\rm Tot}  -  E_A)\, .
	\label{Tot_num}
\end{equation} 
Assuming $W^{-1}(1)=0$ (which is relevant to the stretched exponential case), we have
\begin{equation} \label{S_tot}
	S^{\rm Tot}_\alpha \ = \ W^{-1}\left[\Omega_A(E_A)\Omega_B(E^{\rm Tot}  - E_A)\right]\, .
\end{equation}
The equilibrium distribution of energy between $A$ and $B$ is assumed to correspond to the energies $(E_A$ and $E^{\rm Tot}-E_A)$ that maximize the total number of microstates, i.e., we impose the condition
\begin{equation}
	\frac{\partial S^{\rm Tot}_{\alpha}}{\partial E_A} \ = \ 0\, .\label{Equi_cond}
\end{equation}
According to  Eq. (\ref{S_tot}) we obtain 
\begin{widetext}
\begin{eqnarray}
	&&\frac{\partial S_{\alpha}^{\rm Tot}}{\partial E_A}  \ = \ 	
	\frac{d W^{-1}(t)}{dt}|_{_t=\Omega^{\rm Tot}} \left(\frac{\partial \Omega_A}{\partial E_A}\Omega_B \ - \
	\frac{\partial \Omega_B}{\partial E_B}\Omega_A\right)\nonumber \\[2mm]
	&&\mbox{\hspace{40mm}}\Downarrow \nonumber \\[2mm]
	&&\frac{\partial S_{\alpha}^{\rm Tot}}{\partial E_A} \ = \ \Omega^{\rm Tot}
	\frac{d W^{-1}(t)}{dt}|_{_t=\Omega^{\rm Tot}}\left(\frac{\partial \ln \Omega_A}{\partial E_A} \ - \
	\frac{\partial \ln \Omega_B}{\partial E_B}\right)\nonumber \\[2mm]
	&&\mbox{\hspace{40mm}}\Downarrow \nonumber \\[2mm]
&&\frac{\partial S_{\alpha}^{\rm Tot}}{\partial E_A} \ = \   \Omega^{\rm Tot}
	\frac{d W^{-1}(t)}{dt}|_{_t=\Omega^{\rm Tot}}\left(\frac{\partial \ln W(S_A)}{\partial E_A} \ - \
	\frac{\partial \ln W(S_B)}{\partial E_B}\right).
\end{eqnarray}
\end{widetext}
In the last step, we used Eq.~(\ref{S_uni}) along with the assumption  $W^{-1}(1)=0$. If we further use the fact that
\begin{equation}
	\frac{\partial \ln W(S_{\alpha})}{\partial E} \ = \ \frac{\partial \ln W(t)}{\partial t}|_{_{t=S_{\alpha}}}\frac{\partial S_\alpha}{\partial E}\, ,
\end{equation}
and introduce
\begin{equation}
\frac{\partial S_\alpha}{\partial E} \ \equiv  \ \beta\, ,
    \label{A.68.kl}
\end{equation}
we conclude that the equilibrium condition~(\ref{Equi_cond}) reads
\begin{equation}
	\frac{\partial \ln W(t)}{\partial t}\Big|_{t=S_A}\beta_A \ = \  \frac{\partial \ln W(t)}{\partial t}\Big|_{t=S_B}\beta_B \ = \ \beta^*\, .
    \label{Micro-Equi}
\end{equation}
%
%
This result coincides with the equilibrium condition obtained below in Eq.~(\ref{25cc}) within the framework of macroscopic (i.e., canonical-ensemble-based) thermodynamics. Therefore, the microcanonical and canonical descriptions are consistent at equilibrium and they even share the same empirical temperature; see Eq.~(\ref{T=theta}). As briefly noted in the Introduction, group entropies satisfy the Shore--Johnson axioms and thus lead to unbiased maximum-entropy probability distributions. Consequently, by following Jaynes' construction for the Boltzmann--Gibbs case~\cite{Jaynes1957}, the canonical ensemble associated with group entropies is obtained by maximizing the entropy under the appropriate constraints. This procedure reproduces, under the assumption of ensemble equivalence, the same equilibrium condition as Eq.~(\ref{Micro-Equi}), and thus ensures consistency between the canonical formulation and the microcanonical result.

For some of the most relevant functional forms of $W(N)$, the equilibrium condition 
implied by Eq.~(\ref{Micro-Equi}) becomes
\begin{eqnarray}
	W(N)  &=& N^a 
	\;\;\; \Rightarrow  \;\;\; 
	\frac{\beta_A}{\Omega_A^{1/a}} \ = \ \frac{\beta_B}{\Omega_B^{1/a}}\, ,\\[2mm] 
	W(N)  &=&  k^N 
	\;\;\; \Rightarrow \;\;\; 
	\beta_A \ = \ \beta_B\, , \\[2mm]
	W(N)  &=&   k^{N^\gamma} 
	\nonumber \\[2mm]
    &\Rightarrow& \;\; 
    \frac{\beta_A}{(\ln \Omega_A)^{(1-\gamma)/\gamma}} 
    \ = \ 
    \frac{\beta_B}{(\ln \Omega_B)^{(1-\gamma)/\gamma}}\, .~~~~~
    \label{VI.71.kl}
\end{eqnarray}

\subsection{First law of thermodynamics}

To address the first law of thermodynamics, we recall the basic steps in its derivation in the conventional microcanonical framework. To this end, we loosely follow Reif's elegant derivation from Ref.~\cite{Reif1965}, Sec. 3.8 and 3.9.
We denote the number of accessible states at energy $E$ and with a given set of external parameters $x_1, \ldots, x_m$ as $\Omega(E, x_1, \ldots, x_m)$.
Consider now a quasi-static process in which a system interacts with another system and is taken from an equilibrium state characterized by $\bar{E}$ and $\bar{x}_\nu$ ($\nu = 1,\ldots,m$) to a neighboring equilibrium state specified by $\bar{E} + d\bar{E}$ and $\bar{x}_\nu + d\bar{x}_\nu$. The corresponding change in $\Omega$ is then
 \begin{equation}
	d\Omega \ = \  \frac{\partial \Omega}{\partial E}d\bar{E} \ + \  \sum_{\nu =1}^m \frac{\partial \Omega}{\partial {x}_{\nu}}\ d\bar{x}_\nu\, .
    \label{B.71.cc}
\end{equation}

%

Within the standard Boltzmann--Gibbs microcanonical description, this equation can be transformed into a statement about entropy $S=\ln \Omega(E)$ by simply dividing~(\ref{B.71.cc}) by $\Omega$ and making use of 
\begin{equation}
	\frac{1}{\Omega}\frac{\partial \Omega}{\partial E} \ = \ \frac{\partial \ln \Omega}{\partial E} \ = \ \frac{\partial S}{\partial E} \, .
\end{equation} 
With this, one arrives at
\begin{equation}
	dS \ = \  \frac{1}{T} \Big(dE \ + \ \sum_\nu \bar{X}_\nu d\bar{x}_\nu\Big)\, ,
\end{equation}
where we used relation~(\ref{A.68.kl}) together with the relation
\begin{eqnarray}
\frac{\partial \ln \Omega}{\partial x_{\nu}} \ = \ \frac{1}{T} \bar{X}_{\nu}\, ,
\end{eqnarray}
with $\bar{X}_{\nu}$ representing the {mean generalized force} associated to $x_{\nu}$.

In the more challenging world of group entropies we need to proceed slightly differently. For simplicity, we consider the non-trace form entropy with only a single external parameter $x$, i.e.:
\begin{eqnarray}
&&\mbox{\hspace{-9mm}}S_\alpha \ =  \ \frac{\mathcal{G}\big(\ln \Omega(E)^{1-\alpha}\big)}{1-\alpha}\, .
\end{eqnarray}
This implies that
\begin{eqnarray}
dS_\alpha \ = \ \frac{1}{\Omega}\mathcal{G}'\big(\ln \Omega^{1-\alpha}\big)\left(\frac{\partial \Omega}{\partial E}dE \ + \ \frac{\partial \Omega}{\partial x}dx\right) .
\label{dS}
\end{eqnarray}
We need an estimate of $\partial \Omega/\partial x$. Let $E_r(x)$ denote the energy of microstate $r$ when the external parameter assumes the value $x$. Then 
\begin{equation}
    \Omega(E,x)dE\ = \ |\{r| E_r(x)\in[E,E+dE]\}|\, ,
\end{equation}
and   
\begin{eqnarray}
 &&\mbox{\hspace{-9mm}}\Omega(E,x+dx)dE \ = \ |\{r| E_r(x+dx)\in[E,E+dE]\}|\nonumber\\[2mm]
 &&\mbox{\hspace{-9mm}}= \ |\{r| E_r(x)+\frac{\partial E_r}{\partial x}dx\in[E,E+dE]\}|\nonumber\\[2mm]
 &&\mbox{\hspace{-9mm}}= \ |\{r| E_r(x)\in[E-\frac{\partial E_r}{\partial x}dx,E-\frac{\partial E_r}{\partial x}dx+dE]\}|\, .~~
 \label{VI.80.ck}
\end{eqnarray}
The last expression in Eq.~(\ref{VI.80.ck}) inspires the following expression
\begin{eqnarray}
    &&\mbox{\hspace{-9mm}}\Omega(E,x+dx)dE \ = \ \Omega\Big(E-\Big\langle \frac{\partial E_r}{\partial x}\Big\rangle_{\! r}dx,x\Big)dE\nonumber\\[2mm]   
    &&\mbox{\hspace{5mm}}= \ \Omega(E,x)dE \ - \ \Big\langle \frac{\partial E_r}{\partial x}\Big\rangle_{\!r} \ \! \frac{\partial \Omega}{\partial E}\ \!dxdE\, .
\end{eqnarray}
Here we have introduced the average over all the microstates $r$ indicated by the angular bracket and the subscript $r$. Next, we introduce the generalized force as
\begin{equation}
	\bar{X} \ \equiv \ -\Big\langle \frac{\partial E_r}{\partial x}\Big\rangle_r\, ,
\end{equation}
and  write
\begin{equation}
    \frac{\partial \Omega}{\partial x} \ = \ \bar{X}\ \!\frac{\partial \Omega}{\partial E}\, .
\end{equation}
From Eq.~(\ref{dS}), we can now conclude that
\begin{eqnarray}\label{laws}
	dS_\alpha &=&   \frac{\mathcal{G}'\big(\ln \Omega^{1-\alpha}\big)}{ \Omega}\frac{\partial \Omega}{\partial E}\big(dE \ + \ \bar{X}dx\big)\nonumber \\[2mm]
    &=&\frac{\partial S_{\alpha}}{\partial E}\big(dE \ + \ \bar{X}dx\big)\, .
\end{eqnarray}
As in Eq.~(\ref{A.68.kl}), we can again formally identify   
\begin{equation}
	\beta \ = \ \frac{\partial S_\alpha}{\partial E}\, ,
\end{equation}
and, using Eq.~(\ref{laws}), write 
\begin{equation}
\frac{1}{\beta}dS_\alpha \ = \ dE \ + \ \bar{X}dx\, .
    \label{VI.87.kl}
\end{equation}
Since energy conservation dictates that the right hand side of this equation is equal to the heat exchange, we have 
\begin{equation}
	dS_\alpha \ = \,\beta \,\dbar{Q}\, , 
\end{equation}
consistent with the thermodynamic expressions Eqs.~(\ref{29cf}) and (\ref{heat}) derived in the next sections.

In this section, we have shown that within the group-entropic framework the zeroth and first laws of thermodynamics can be derived directly from the microcanonical ensemble. It should, however, be kept in mind that the equivalence between the microcanonical and canonical ensembles is not guaranteed in general. For systems with long-range interactions, such as gravity, this equivalence may be violated;  see, e.g., Refs.~\cite{Rufo:14,LB,Thirring}. We do not consider the canonical ensemble here; instead, we analyse macroscopic thermodynamics in the following Secs.~\ref{Sec.III.aaa} and~\ref{second-law}.

\section{Zeroth law of thermodynamics: empirical temperature and pressure \label{Sec.III.aaa}}

We now turn to group entropies and macroscopic thermodynamics. Thus, we neglect the underlying microstates and examine whether the algebraic structure of the group entropies allows them to be regarded as state functions satisfying the laws of thermodynamics.

For simplicity, we denote by $S_{\alpha}$ a generic representative of the family of group entropies defined in Eq.~\eqref{eq:Z}. First, we consider the zeroth law of thermodynamics, which formalizes the concept of thermal equilibrium~\cite{Huang,Reif1965,L-L:2013}. It states that if two thermodynamic systems are in thermal and mechanical equilibrium with each other and each is separately in equilibrium with a third system, then all three systems are mutually in thermal equilibrium. This transitive property allows all systems to be grouped into equivalence classes, each with unique labels (or quantifiers) known as the empirical temperature and physical pressure. It should be stressed that an empirical temperature scale does not necessarily coincide with the absolute temperature; the latter is a consequence of the second law. In general, the empirical temperature depends on the thermometer's construction and on the medium with which it exchanges heat, whereas the absolute temperature remains unaffected by these factors and reflects only the intrinsic thermodynamic state of the system.
For example, in conventional thermodynamics, the Celsius and Fahrenheit scales serve as empirical temperature scales, whereas the Kelvin and Rankine scales represent the corresponding absolute temperature scales.

Although the zeroth law of thermodynamics does not, in principle, require the concept of entropy --- since entropy arises naturally from the second law --- its explicit use can be advantageous, as it allows empirical temperature and physical pressure to be formulated in a more quantitative manner~\cite{L-L:2013}. 


We will assume that while the group entropy is generally non-additive  (but extensive), the energy remains additive.  This \textit{modus operandi} is often employed in the study of systems with long-range interactions or strong correlations~\cite{Abe,Biro}, and it also arises naturally in scenarios where statistical correlations lead to non-additive entropy. Although this framework is admittedly restrictive, it represents a phenomenologically useful starting point that allows for a controlled analysis.

With this in mind, we consider two systems, $A$ and $B$, allowing only weak interactions between them in order to establish both thermal and mechanical contact. Note that the interactions within $A$ and within $B$ may still be long-range and strong. The volumes of the subsystems  are denoted by $V(A)$ and $V(B)$, and their internal energies by $U(A)$ and $U(B)$.  For the composite system, we assume conservation of total volume and total internal energy. Hence, we assume that the allowed states of the combined system $A\times B$ consist of the Cartesian combination of internal states of $A$ with internal states of $B$. That is, we do not allow new states which might correspond to interactions between components from $A$ and components from $B$. Hence we have 
\begin{eqnarray}
&&U(A\times B)  \ = \ U(A) \ + \ U(B) \ = \  U_{\text{tot}}  \ = \ {const.}\, , 
\nonumber \\[2mm] 
&&V(A\times B)  \ = \ V(A) \ + \  V(B) = V_{\text{tot}} \ = \ {const.} \, .~~~~~~~~
\label{23cf}
\end{eqnarray}
We assume that at thermodynamic equilibrium, the composite entropy 
$S(A\times B)$, attains its maximum value subject to the constraints of fixed $U_{\text{tot}}$, $V_{\text{tot}}$, etc.. 
This maximization principle ensures that no further spontaneous redistribution of energy or matter between subsystems can occur. 
%
Taking into account that $\Phi(x,y)=G(G^{-1}(x)+G^{-1}(y))$, we can write
\beq
S_{\alpha}(A \times B) \ = \   G\big(G^{-1}(S_{\alpha}(A)) \ + \ G^{-1}(S_{\alpha}(B))\big)\, , 
\label{Group_compo}
\eeq
and thus, we deduce
%
%
\begin{widetext}
\begin{eqnarray} \label{34.jk}
&&\mbox{\hspace{-4mm}}0 \  = \ {d}S_{\alpha}(A \times B) \nonumber \\[2mm]
&&\mbox{\hspace{-4mm}}= G'\big(G^{-1}(S_{\alpha}(A))+G^{-1}(S_{\alpha}(B))\big)\left\{\Big[\frac{d G^{-1}(t)}{dt}\Big|_{t=S_{\alpha}(A)}\left(\frac{\partial S_{\alpha}(A)}{\partial U(A)}\right)_{V(A)} - \frac{d G^{-1}(t)}{dt}\Big|_{t=S_{\alpha}(B)}\left(\frac{\partial S_{\alpha}(B)}{\partial U(B)}\right)_{V(B)}\Big]\ \!{d}U(A) \right.\nonumber \\[2mm]
&&\mbox{\hspace{-4mm}}+  \left.
\Big[\frac{d G^{-1}(t)}{dt}\Big|_{t=S_{\alpha}(A)}\left(\frac{\partial S_{\alpha}(A)}{\partial V(A)}\right)_{U(A)}  - \ \frac{d G^{-1}(t)}{dt}\Big|_{t=S_{\alpha}(B)}\left(\frac{\partial S_{\alpha}(B)}{\partial V(B)}\right)_{U(B)}\Big]\ \! {d}V(A)\right\} \, .
\end{eqnarray}
\end{widetext}

Here we have assumed that the group entropy $S_{\alpha}$ is expressed in terms of its natural state variables, namely $U$ and $V$. However, the analysis can also be carried out using other, non-mechanical natural variables, such as particle number, magnetization, polarization, and so on.

Now, we can relate the factor $	{d G^{-1}(t)}/{d t}$ to the state space growth rate function $W(N)$. This is due to the fact that~\cite{J-Temp,TJ2020SR}
\begin{equation}
	G(t) \ = \  \lambda \Big[W^{-1}(\exp t) \ - \ W^{-1}(1)\Big]\, .
\end{equation}
For simplicity, let us assume that $\lambda = 1$ and that $W^{-1}(1) = 0$. This holds, for instance, for the stretched exponential entropy. In particular, the condition $W^{-1}(1) = 0$ is not an essential restriction. As in conventional statistical thermodynamics, the limit $N \to \infty$ should be taken only at the very end, after all intermediate manipulations have been completed. In this framework, any contribution that is universal --- such as the term $W^{-1}(1)$ --- does not affect the thermodynamic limit. Consequently, different asymptotic behaviours of $W(N)$ should lead to the same form of Eq.~(\ref{Micro-Equi}) as $N \to \infty$.
Therefore, the universal term $W^{-1}(1)$ could potentially even be removed from the definition of the entropy, in analogy with conventional thermodynamics, where one subtracts the universal (state independent) zero-temperature contribution to the entropy (leading to Planck's formulation of the third law). 
Then,
\begin{equation}
	G(t) \ = \  W^{-1}(\exp t) \;\rightarrow \;  G^{-1}(t) \ = \  \ln[W(t)]\, ,
\end{equation}
and 
\begin{equation} \label{eq:20}
	\frac{d G^{-1}(t)}{d t} \ = \  \frac{W'(t)}{W(t)}\, .
\end{equation}
From~\eqref{34.jk}, we can derive  two identities that reflect the simultaneous thermal and mechanical equilibrium of systems $A$ and $B$.
The {\em first} identity can be expressed as 
%
%
%
\begin{eqnarray}
	&&\mbox{\hspace{-6mm}}\frac{W'(S_{\alpha}(A))}{W(S_{\alpha}(A))}\left(\frac{\partial S_{\alpha}(A)}{\partial U(A)}\right)_{V(A)} \nonumber \\[2mm] &&\mbox{\hspace{6mm}}= \ \frac{W'(S_{\alpha}(B))}{W(S_{\alpha}(B))}\left(\frac{\partial S_{\alpha}(B)}{\partial U(B)}\right)_{V(B)}\, .
    \label{Equi_con}
\end{eqnarray}
If, by analogy with conventional thermodynamics, we formally define
\begin{equation}\label{I.15.kl}
\beta \ \equiv \ \left(\frac{\partial S_{\alpha}}{\partial U}\right)_{V}\,,
\end{equation} 
we obtain the  relation
%
\begin{eqnarray} \label{25cc}
 \beta(A) \frac{W'(S_{\alpha}(A))}{W(S_{\alpha}(A))} \ &=& \ \beta(B)\frac{W'(S_{\alpha}(B))}{W(S_{\alpha}(B))}\nonumber \\[2mm]&\equiv& \  \beta^*\, .
\end{eqnarray}
This expression is identical to the equilibrium condition, Eq. (\ref{Micro-Equi}), derived within the microcanonical ensemble.

It should be stressed that the true empirical temperature (or physical temperature) does not coincide with the formal expression  $( \beta)^{-1}$. Instead, it is determined via the relation
%
\begin{eqnarray}
\vartheta \ \equiv \ \frac{1}{\beta^*} \ = \ \frac{W(S_{\alpha})} {W'(S_{\alpha})}\frac{1}{ \beta}\, .
\label{29cf}
\end{eqnarray}
Eqs.~\eqref{25cc} and~\eqref{29cf} encapsulate the zeroth law of thermodynamics, which guarantees that one can assign the same {\em empirical} temperature $\vartheta$ to all subsystems in {\em thermal} equilibrium. In our case, this implies $\vartheta_A = \vartheta_B$. 
We note that, in general, the parameter $\vartheta$ is a function of both $\alpha$ and $\gamma$; however, for brevity, this dependence will be suppressed in the subsequent analysis.
At this stage, the temperature is only empirical because the absolute temperature is a logical consequence of the second law of thermodynamics and not of the zeroth law. 
However, it is clear that, since the concept of entropy was used, the empirical temperature Eq.~(\ref{25cc}) will not be entirely blind to the existence of absolute temperature. This point will be discussed in more detail in Section~\ref{IV.cf}.

The {\em second} identity that follows from Eqs.~(\ref{34.jk}) and \eqref{eq:20} can be written as
%
%
\begin{eqnarray}
&&\mbox{\hspace{-5mm}}{\left(\frac{\partial S_{\alpha}(A)}{\partial V(A)}\right)_{\!U(A)}  } \ \!\frac{W'(S_{\alpha}(A))}{W(S_{\alpha}(A))}  \nonumber \\[2mm] 
&&= \  {\left(\frac{\partial S_{\alpha}(B)}{\partial V(B)}\right)_{\!U(B)}  } \ \!\frac{W'(S_{\alpha}(B))}{W(S_{\alpha}(B))}
\ \equiv\ \frac{p_{\rm{phys}}}{\vartheta}\, .~
\label{29ccg}
\end{eqnarray}
As in the case of empirical temperature, $p_{\rm{phys}}$ is generally dependent on the same parameters as the considered group entropy $S_{\alpha}$. 

Eq.~(\ref{29ccg}) reflects that when two systems are in mechanical equilibrium, their pressures are equal. This allows us to identify {\em physical pressure}, $p_{\rm{phys}}$, as
%
\begin{eqnarray}
p_{\rm{phys}} \ = \ {\vartheta}\ \!\frac{W'(S_{\alpha})}{W(S_{\alpha})} \left(\frac{\partial{S_{\alpha}}}{\partial V}\right)_{\!U}\, .
\label{31bb}
\end{eqnarray}
When the number of microstates scales exponentially with $N$, as in standard thermodynamics, one obtains $W'(S_{\alpha}
)/W(S_{\alpha}) = 1$. Consequently, in such a case Eq.~(\ref{31bb}) reproduces the familiar thermodynamic result with an intensive empirical temperature.

\section{Carath\'{e}odory's principle and absolute temperature \label{second-law}}

Let us now turn to the second law of thermodynamics to examine how the 
absolute temperature, if it exists, relates to the empirical temperature. 
For non-additive entropies, such as group entropies, the situation is far less 
straightforward than in standard thermodynamics. In particular, it is not 
\emph{a priori} evident what the conjugate thermodynamic variable should be 
for a group entropy, nor is it clear whether a group entropy defines a proper 
state function.  

In the context of classical,  Clausius-type thermodynamics, Carath\'{e}odory's formulation 
of the second law addresses both of these 
issues simultaneously~\cite{Caratheodory,CaratheodoryII}. Specifically, it guarantees that the heat one-form 
(or Pfaffian form) $\dbar Q$ admits an integrating factor, i.e., that the heat 
one-form is {\em holonomic}. In turn, the resulting exact differential then defines a new 
state function --- the entropy. 
In the case of additive entropies, the inverse of the integrating factor can be equated with the absolute temperature by invoking the Carnot cycle argument and the related Clausius equality~\cite{Huang}. 
However, since $S_{\alpha}$ is not generally additive, the Carnot argument breaks down. Consequently, it is no longer justified to simply equate the integrating factor with the inverse absolute temperature.
To make this fact clearer, we examine the structure of the exact differential associated with the heat one-form  $\dbar {{Q}}$. In Carath\'{e}odory's formulation of the second law, this differential defines a new state function, namely the entropy. It is therefore natural to ask whether an integrating factor exists such that 
\begin{eqnarray}
dS_{\alpha}({\pmb{\xi}},\vartheta) \ = \ \mu({\pmb{\xi}}, \vartheta) \  \dbar {{Q}}({\pmb{\xi}},\vartheta)\, , 
\end{eqnarray}
where $\pmb{\xi}$ represents a collection of relevant state variables (such as volume $V$, magnetization ${\mathbf{M}}$, number of particles $N$, etc.) and $\vartheta$ is some {\em empirical} temperature whose existence is guaranteed by the zeroth law of thermodynamics (see Section \ref{Sec.III.aaa}).

Let us now divide the thermodynamic system under consideration into two subsystems $A$ and $B$ that are described by the state variables
$\{{\pmb{\xi}}_1,\vartheta\}$ and $\{{\pmb{\xi}}_2,\vartheta\}$, respectively. Then
\begin{eqnarray}
&&\dbar {{Q}}_A({\pmb{\xi}}_1,\vartheta) \ = \ \frac{1}{\mu_A({\pmb{\xi}}_1,\vartheta)} \ \! dS_{\alpha}(A)({\pmb{\xi}}_1,\vartheta)\, , \nonumber \\[2mm]
&&\dbar {{Q}}_B({\pmb{\xi}}_2,\vartheta) \ = \ \frac{1}{\mu_B({\pmb{\xi}}_2,\vartheta)} \ \! dS_{\alpha}(B)({\pmb{\xi}}_2,\vartheta)\, .
\end{eqnarray}
Thus, for the whole system, we have
\begin{eqnarray}
\dbar {{Q}}_{A \times B} \ = \ \ \dbar {{Q}}_A \ + \ \ \dbar {{Q}}_B\, ,
\end{eqnarray}
with 
\begin{widetext}
\begin{eqnarray}
\dbar {{Q}}_{A\times B}({\pmb{\xi}}_1,{\pmb{\xi}}_2, \vartheta)  \ = \ \frac{1}{\mu_{A \times B}({\pmb{\xi}}_1,{\pmb{\xi}}_2,\vartheta)} \ \! dS_{\alpha}(A\times B)({\pmb{\xi}}_1,{\pmb{\xi}}_2,\vartheta)\, .
\end{eqnarray}
Thus, we can write
\begin{eqnarray}
dS_{\alpha}(A\times B)({\pmb{\xi}}_1,{\pmb{\xi}}_2,\vartheta) \ = \ \frac{\mu_{A \times B}({\pmb{\xi}}_1,{\pmb{\xi}}_2,\vartheta)}{\mu_A({\pmb{\xi}}_1,\vartheta)} \ \! dS_{\alpha}(A)({\pmb{\xi}}_1,\vartheta) 
 \ + \
 \frac{\mu_{A\times B}({\pmb{\xi}}_1,{\pmb{\xi}}_2,\vartheta)}{\mu_B({\pmb{\xi}}_2,\vartheta)} \ \! dS_{\alpha}(B)({\pmb{\xi}}_2,\vartheta)\, .
 \label{B.14.fg}
\end{eqnarray}
We now assume, for simplicity, that there is only one state variable (e.g., volume $V$) in addition to temperature, so that ${\pmb{\xi}} = \xi$. If there were more state variables, our subsequent argument would still be valid, but we would need to consider more than two subsystems. Under this simplifying assumption, the entropies $S_{\alpha}(A)(\xi_1,\vartheta)$ and $S_{\alpha}(B)(\xi_2,\vartheta)$ can be locally inverted with respect to $\xi_1$ and $\xi_2$. Consequently, we may write
\begin{eqnarray}
\xi_1 \ = \ \xi_1(S_{\alpha}(A), \vartheta) \;\;\; \mbox{and} \;\;\; \xi_2 \ = \ \xi_2(S_{\alpha}(B), \vartheta) \, .
\end{eqnarray}
With this, we can cast~(\ref{B.14.fg}) into the following form:
\begin{eqnarray}
dS_{\alpha}(A\times B)(S_{\alpha}(A),S_{\alpha}(B),\vartheta) \ &=& \ \frac{\mu_{A \times B}(S_{\alpha}(A),S_{\alpha}(B),\vartheta)}{\mu_A(S_{\alpha}(A),\vartheta)} \ \! dS_{\alpha}(A) \nonumber \\[2mm]
&+& \
 \frac{\mu_{A\times B}(S_{\alpha}(A),S_{\alpha}(B),\vartheta)}{\mu_B(S_{\alpha}(B),\vartheta)} \ \! dS_{\alpha}(B) \ + \ 0 \ \! d\vartheta\, .
 \label{B.16.cb}
\end{eqnarray}
Since we require that $dS_{\alpha}$ be an exact differential, one gets the following integrability conditions:
\begin{eqnarray}
&&\mbox{\hspace{-10mm}}\frac{\partial \log\big(\mu_A(S_{\alpha}(A),\vartheta)\big)}{\partial \vartheta} \ = \ \frac{\partial \log\big(\mu_B(S_{\alpha}(B),\vartheta)\big)}{\partial \vartheta} \ = \ \frac{\partial \log\big(\mu_{A \times B}(S_{\alpha}(A),S_{\alpha}(B),\vartheta)\big)}{\partial \vartheta}\, , \label{B.17a.cv} \\[2mm]
&&\mbox{\hspace{-10mm}} \frac{1}{\mu_A(S_{\alpha}(A),\vartheta)} \frac{\partial \mu_{A\times B}(S_{\alpha}(A),S_{\alpha}(B),\vartheta)}{\partial S_{\alpha}(B)} \ = \ \frac{1}{\mu_B(S_{\alpha}(B),\vartheta)} \frac{\partial \mu_{A \times B}(S_{\alpha}(A),S_{\alpha}(B),\vartheta)}{\partial S_{\alpha}(A)}\, . \label{B.17b.cv}
\end{eqnarray}
If all the integrability conditions can be satisfied, then, $S_{\alpha}$ is a state function. 

Let us first focus on Eq.~(\ref{B.17a.cv}). From its structure, we see that the derivatives involved cannot depend on the entropy, but only on $\vartheta$ and on possible indices specific to the considered entropy. Thus, we can denote the right-hand side of~(\ref{B.17a.cv}) generically as $-m_{\alpha}(\vartheta)$ (where $m_{\alpha}$ is some function of $\vartheta$ and the minus sign is introduced merely for future convenience). The solutions can be readily obtained by separation of variables, namely
\begin{eqnarray}
&&\mu_A\big(S_{\alpha}(A),\vartheta\big ) \ = \ \Gamma_A\big(S_{\alpha}(A)\big) \ \!\exp\left(-\int m_{\alpha}(\vartheta) d\vartheta \right) \ = \ \Gamma_A\big(S_{\alpha}(A)\big) \ \! T^{-1}_{\alpha}(\vartheta)\, ,\nonumber \\[2mm]
&&\mu_B(S_{\alpha}(B),\vartheta) \ = \ \Gamma_B\big(S_{\alpha}(B)\big) \ \!\exp\left(-\int m_{\alpha}(\vartheta) d\vartheta \right)  \ = \ \Gamma_B\big(S_{\alpha}(B)\big) \ \! T^{-1}_{\alpha}(\vartheta)\, ,\nonumber \\[2mm]
&&\mu_{A\times B}(S_{\alpha}(A),S_{\alpha}(B),\vartheta) \ = \ \Gamma_{A\times B}\big(S_{\alpha}(A),S_{\alpha}(B)\big) \ \!\exp\left(-\int m_{\alpha}(\vartheta) d\vartheta \right)  \ = \ \Gamma_{A\times B}\big(S_{\alpha}(A),S_{\alpha}(B)\big)  \ \! T^{-1}_{\alpha}(\vartheta)\,.~~~~~~~~~
\label{B.19.hh}
\end{eqnarray}
\end{widetext}
Here $\Gamma_{X}$ ($X$ stands for $A$, $B$ and $A\times B$, respectively) are some arbitrary functions of the entropy and $T_{\alpha}(\vartheta)$,  given by 
\begin{equation}
    T_{\alpha}(\vartheta) \ \equiv \ \exp\left(\int m_{\alpha}(\vartheta) d\vartheta \right),
    \label{Def_temp}
\end{equation}
is a subsystem-independent  function of the empirical temperature. The sign in front of $m_{\alpha}(\vartheta)$ was adopted to ensure that the temperature function $T_{\alpha}(\vartheta)$ is a monotonically increasing function of the empirical temperature, given that $m_{\alpha}(\vartheta)$ is a positive function. In conventional thermodynamics this is always the case~\cite{L-L:2013, Huang}.

Thus, we have succeeded in finding integrating factors.
In the theory of Pfaffian forms it is well-known that if one integrating factor $\mu$ exists, i.e., the equation 
\begin{eqnarray}
dS_{\alpha}({\pmb{\xi}},\vartheta) \ = \ \mu({\pmb{\xi}}, \vartheta) \  \dbar {{Q}}({\pmb{\xi}},\vartheta)\, ,
\end{eqnarray}
is fulfilled, then any other function $\tilde{\mu} = f(S_{\alpha})\mu$
(where $f$ is an arbitrary positive, locally $L^1$ function) is again an integrating factor because
\begin{eqnarray}
\tilde{\mu} \  \dbar {{Q}} \ = \ f(S_{\alpha}) dS_{\alpha} \ = \  d \left( \int^{S_{\alpha}}_a f(s) ds\right) \ \equiv \ d \tilde{S}_{\alpha}\, ,~~~
\end{eqnarray}
($a$ is an arbitrary integration constant).  In classical thermodynamics, one selects from all possible integrating factors only those that depend on the empirical temperature $\vartheta$, and not on $S_{\alpha}$. This requirement stems, on the one hand, from Carnot's theorem, which states that the absolute temperature depends solely on the empirical temperature and not on any other state variable, and, on the other hand, from the Clausius equality, which identifies the absolute temperature with the inverse of the integrating factor (see, e.g., Ref.~\cite{L-L:2013}). Accordingly, in classical thermodynamics, $\Gamma_X$ is taken to be constant --- essentially fixing the choice of units --- thereby allowing the integrating factor to be expressed as
\begin{eqnarray}
\exp\!\left(-\int m_{\alpha}(\vartheta)\, d\vartheta \right).
\label{II.35.cc}
\end{eqnarray}
It is straightforward to verify that the integrating factor~(\ref{II.35.cc}) also satisfies the remaining integrability condition~(\ref{B.17b.cv}).

In the framework of group entropies, neither Carnot's theorem nor the Clausius equality is available to guide the selection of the integrating factor. Fortunately, the non-additive nature of group entropies allows the integrating factor to be determined uniquely (up to a multiplicative constant). Indeed, by differentiating Eq.~(\ref{Group_compo}), we find that
%
%
%
\begin{widetext}
\begin{eqnarray}
dS_{\alpha}(A \times  B) \ = \  G'\Big(G^{-1}(S_{\alpha}(A)) \ + \ G^{-1}(S_{\alpha}(B))\Big)\Big[\frac{d G^{-1}(t)}{dt}\Big|_{t=S_{\alpha}(A)}\!\!\!\!\,\, \, dS_{\alpha}(A) \ + \ \frac{d G^{-1}(t)}{dt}\Big|_{t=S_{\alpha}(B)}\!\!\!\! \, \, \, dS_{\alpha}(B)\Big]. \nonumber \\
\end{eqnarray}
\end{widetext}
Due to Eq.~\eqref{eq:20}, we obtain the general relation
%
\[
\frac{d G^{-1}(t)}{dt}\Big|_{t=S_{\alpha}} \ = \  \frac{W'(S_{\alpha})}{W(S_{\alpha})}\, .
\]
By comparing this with Eqs.~(\ref{B.16.cb}) and~(\ref{B.19.hh}), we obtain after some algebraic manipulations
%
%
\begin{widetext}
\begin{eqnarray}
\Gamma_{A}\frac{W'(S_{\alpha}(A))}{W(S_{\alpha}(A))} \ = \ \Gamma_{B}\frac{W'(S_{\alpha}(B))}{W(S_{\alpha}(B))} \ = \ \Gamma_{A\times B} \frac{W'(S_{\alpha}(A \times B))}{W(S_{\alpha}(A\times  B))} \ \equiv \ \eta \, ,
\label{Theo_Equi}
\end{eqnarray}
\end{widetext}
where $\eta $ is an arbitrary constant. Thus, we can infer that the condition 
%
\beq
\Gamma_X(S_{\alpha}(X)) \ = \  \eta  \ \! \frac{W(S_{\alpha}(X))}{W'(S_{\alpha}(X))}\, ,
\label{Omega_W}
\eeq
must hold.  

We thus see that it is not possible to set the value of  $\Gamma(S_{\alpha})$ to a constant, unless the considered group entropy $S_{\alpha}$ coincides with R\'enyi's entropy, i.e., $W(N)\sim \exp N $. Consequently, the integrating factor cannot be chosen to be merely a function dependent on $\vartheta$ (as in classical thermodynamics).    However, since an integrating factor does exist for $S_{\alpha}$  [indeed, one can verify that~(\ref{B.19.hh}), together with the above $\Gamma$, 
also satisfies the second integrability condition~(\ref{B.17b.cv})], 
the group entropy $S_{\alpha}$ may be regarded as a genuine state function --- though with the important distinction that this integrating factor can no longer be identified with the inverse absolute temperature. 

\section{Empirical vs. absolute temperature in non-additive thermodynamics\label{IV.cf}}

Let us return to Sec.~\ref{Sec.III.aaa} and assume that our group entropy is expressed
in terms of its natural state variables $U$ and $V$ (more could be easily added),  i.e., $S_{\alpha} = S_{\alpha}(U,V)$. Since $S_{\alpha}$ is a state function, its infinitesimal change resulting from $U\mapsto U + dU$ and $V \mapsto V+ dV$ is a total differential, i.e.,
%
\begin{eqnarray}
dS_{\alpha} &=& \ \left(\frac{\partial S_{\alpha}}{\partial U}\right)_{\!V} dU 
 \ + \  \left(\frac{\partial S_{\alpha}}{\partial V}\right)_{\!U}dV \nonumber \\[2mm]
&=& \frac{W(S_{\alpha})}{W'(S_{\alpha})} \ \!\frac{1}{\vartheta} \ \! dU \ 
 + \frac{W(S_{\alpha})}{W'(S_{\alpha})} \!\frac{p_{\rm{phys}}}{\vartheta} \ \! dV  \, ,~~~~~~~
 \label{IV.33.kl}
\end{eqnarray}
where in the second equality we used Eqs.~(\ref{I.15.kl}), (\ref{29cf}) and~(\ref{31bb}). We can equivalently rewrite Eq.~(\ref{IV.33.kl})  as
%
%
\begin{eqnarray} \label{eq:46}
\frac{W'(S_{\alpha})}{W(S_{\alpha})} \ \! \vartheta dS_{\alpha} \ = \ dU 
 \ + \ p_{\rm{phys}} \ \!dV\, .
\end{eqnarray}
According to the first law of thermodynamics (i.e., energy conservation), 
the right-hand side can be identified with the heat one-form, $\dbar Q$; 
consequently, we may write [cf. also Eq.~(\ref{29cf})],
%
%
\begin{eqnarray}\label{heat}
&& \dbar {{Q}} \ = \ \frac{W'(S_{\alpha})}{W(S_{\alpha})}  \vartheta dS_{\alpha} \ = \  \frac{1}{ \beta}  dS_{\alpha}\, .
\end{eqnarray}
On the other hand, from the previous section we know that [see Eqs.~(\ref{B.19.hh}) and (\ref{Omega_W})] 
%
%
\begin{eqnarray} \label{eq:48}
 \dbar {{Q}} \ = \ \frac{T}{\Gamma} \, dS_{\alpha} \ = \ \frac{W'(S_{\alpha})}{W(S_{\alpha})} \frac{T} {\eta } \ \! d S_{\alpha}\, ,
 \label{IV.36.c}
\end{eqnarray}
(the prospective index $\alpha$ is again suppressed  in both $\vartheta$ and $T$).
This implies the equality 
\begin{eqnarray}
T(\vartheta) \ = \  \eta  \vartheta\, .
\label{T=theta}
\end{eqnarray}
In other words, the empirical temperature $\vartheta$  introduced in Eq.~(\ref{29cf}) coincides with the function $T$ introduced in Eq.~(\ref{Def_temp})  when expressed in the appropriate units. This also implies that $m(\vartheta) = 1/\vartheta$, which is indeed a positive function. 

To broaden our insight, let us ignore Eq.~(\ref{T=theta}), or any explicit functional relation between $T$ and  $\vartheta$, and focus solely on relation~(\ref{IV.36.c}) 
to show that the thermal equilibrium condition~\eqref{25cc} can be phrased in the form 
\begin{eqnarray}
T_A \ = \ T_B\, , 
\end{eqnarray}
which also holds in conventional thermodynamics. One might naturally wonder whether $T(\vartheta)$ could serve as a universal absolute temperature, even though it is not a thermodynamic variable conjugate to the entropy $S_{\alpha}$. We now show that this is indeed the case.

Thus, we are interested in how $T$ transforms when moving from one empirical temperature scale $\{\vartheta_1\}$, to another, $\{\vartheta_2\}$. In particular, Eq.~(\ref{eq:48}) 
allows us to write
%
%
%
\begin{eqnarray}
&&\mbox{\hspace{-5mm}} \frac{1}{\eta }\frac{W'(S_{\alpha})}{W(S_{\alpha})} \, dS_{\alpha} \ = \  \frac{1}{\eta }                 \, d \, [\log W(S_{\alpha})] \nonumber \\[2mm] &&\mbox{\hspace{-2mm}}= \ \frac{1}{T} \left\{\left(\frac{\partial U}{\partial T}  \right)_{\!V} dT +  \left[ \left(\frac{\partial U}{\partial V} \right)_{\!T} +   p_{\rm{phys}}  \right] dV   \right\}.~~~~~
\end{eqnarray}
Since the left-hand side is a total differential, the coefficient functions of $dT$ and $dV$ must satisfy the integrability condition
\begin{eqnarray}
&&\frac{\partial }{\partial T} \left\{ \frac{1}{T} \left[ \left(\frac{\partial U}{\partial V} \right)_{\!T} \ + \  p_{\rm{phys}}  \right] \right\}_{V} \nonumber \\[2mm] &&~~~~~~~~~~~~~~~~~~~~~~ =\ \frac{\partial}{\partial V} \left\{ \frac{1}{T} \left(\frac{\partial U}{\partial T} \right)_{\!V} \right\}_{\!T}\,. 
\end{eqnarray}
This condition implies the relation
\begin{eqnarray}
\left(\frac{\partial U}{\partial V} \right)_{\!T} \ + \  p_{\rm{phys}}  \ = \ T \left(\frac{\partial p_{\rm{phys}}}{\partial T} \right)_{\!V}\, .
\label{IV.40.cc}
\end{eqnarray}
Since $T = T(\vartheta)$, it follows that, for $T= \mathrm{const.}$, also $\vartheta = \mathrm{const.}$, and thus 
\begin{eqnarray}
\left(\frac{\partial U}{\partial V} \right)_{\!T}  \ = \ \left(\frac{\partial U}{\partial V} \right)_{\!\vartheta} \, .
\end{eqnarray}
With this we can rewrite Eq.~(\ref{IV.40.cc}) in terms of $\vartheta$ as
\begin{eqnarray}
\left(\frac{\partial U}{\partial V} \right)_{\!\vartheta} \ + \  p_{\rm{phys}}  \ = \ T \left(\frac{\partial p_{\rm{phys}}}{\partial \vartheta} \right)_{\!V} \left(\frac{d T(\vartheta)}{d \vartheta}\right)^{-1}. ~~~~
\label{IV.42.hj}
\end{eqnarray}
Thus, from Eqs.~(\ref{Def_temp}) and~(\ref{IV.42.hj}), we have
\begin{eqnarray}
\frac{d \log T(\vartheta) }{ d\vartheta} \ = \  \frac{\left(\frac{\partial p_{\rm{phys}}}{\partial \vartheta} \right)_{\!V} }{\left(\frac{\partial U}{\partial V} \right)_{\!\vartheta} \ + \  p_{\rm{phys}}} \ = \ m(\vartheta)\, ,
\label{IV.43.kl}
\end{eqnarray}\\
(the prospective index $\alpha$ is suppressed in  $m_{\alpha}$, $\vartheta_{\alpha}$ and $T_{\alpha}$). This demonstrates that, in principle, $m(\vartheta)$ can be experimentally determined for any empirical temperature. Now, let us see that the $\{T\}$ scale  is independent of the chosen empirical ``thermometer'' ---  i.e., it  can be identified with an absolute temperature scale.
To see this, we will assume that the same equilibrium state is described by two different empirical thermometers with scales  $\{\vartheta_1\}$ and $\{\vartheta_2\}$. We further assume that the two scales are monotonic functions of each other, so that two different empirical temperatures on one scale cannot correspond to the same temperature on the other (otherwise such empirical thermometers would not be admissible). Thus, we have $\vartheta_1 = \vartheta_1(\vartheta_2)$ and we let $T_1 = T_1(\vartheta_1)$ and $T_2 = T_2(\vartheta_2)$ denote the two temperature scales $\{T\}$ derived from the respective empirical temperature scales. From Eq.~(\ref{IV.43.kl}), we have
\begin{widetext}
\begin{eqnarray}
&&\mbox{\hspace{-5mm}}\frac{d \log T_1(\vartheta_1) }{ d\vartheta_1} \ = \  \frac{\left(\frac{\partial p_{\rm{phys}}}{\partial \vartheta_1} \right)_{\!V} }{\left(\frac{\partial U}{\partial V} \right)_{\!\vartheta_1} \ + \  p_{\rm{phys}}} \ = \ m_1(\vartheta_1)\, , \nonumber \\[2mm]
&&\mbox{\hspace{-5mm}}\frac{d \log T_2(\vartheta_2) }{ d\vartheta_2} \ = \  \frac{\left(\frac{\partial p_{\rm{phys}}}{\partial \vartheta_2} \right)_{\!V} }{\left(\frac{\partial U}{\partial V} \right)_{\!\vartheta_2} \ + \  p_{\rm{phys}}} \ = \ m_2(\vartheta_2)\ = \ \frac{\left(\frac{\partial p_{\rm{phys}}}{\partial \vartheta_1} \right)_{\!V} \frac{d\vartheta_1}{d \vartheta_2} }{\left(\frac{\partial U}{\partial V} \right)_{\!\vartheta_1} \ + \  p_{\rm{phys}}} \ = \ m_1(\vartheta_1) \frac{d\vartheta_1}{d \vartheta_2} \, .
\end{eqnarray}
From this, we can determine the relationship between the temperature scales  $\{T_1\}$ and $\{T_2\}$. In particular [cf. the defining relation~(\ref{Def_temp})]
\begin{eqnarray}
T_2(\vartheta_2) \ = \ T_2(\vartheta_{02}) \exp\left( \int_{\vartheta_{02}}^{\vartheta_2} m_2(\vartheta_2') d\vartheta_2' \right) \ = \  T_2(\vartheta_{02}) \exp\left( \int_{\vartheta_{02}}^{\vartheta_2} m_1(\vartheta_1') \frac{d\vartheta_1'}{d\vartheta_2'} d\vartheta_2' \right) \ = \ \frac{T_2(\vartheta_{02})}{T_1(\vartheta_{01})} \ \! T_1(\vartheta_1)\, .
\label{IV.45.cf}
\end{eqnarray}
\end{widetext}
Here $\vartheta_{01}$ and $\vartheta_{02}$ are  empirical temperatures that describe the same reference physical state.
Thus, we obtain
\begin{eqnarray}
\frac{T_2(\vartheta_2)}{T_2(\vartheta_{02})} \ = \ \frac{T_1(\vartheta_1)}{T_1(\vartheta_{01})}\, .
\label{IV.46.kl}
\end{eqnarray}
At this stage, we may assume that a generic equilibrium state is characterized by the temperatures $\vartheta_1^*$ and $\vartheta_2^*$. If we choose the units of the temperature scales $\{T_1\}$ and $\{T_2\}$ so that the temperature difference between two states --- described either by the empirical temperatures $\vartheta_{01}$ and $\vartheta_1^*$ or equivalently by $\vartheta_{02}$ and $\vartheta_2^*$ --- is the same, i.e.,
\begin{eqnarray}
T_1(\vartheta_1^*) \ - \ T_1(\vartheta_{01}) \ = \ T_2(\vartheta_2^*) \ - \ T_2(\vartheta_{02})\, ,
\label{V.60.kk}
\end{eqnarray}
then, this together with~(\ref{IV.46.kl}) directly implies that  
\begin{eqnarray}
T_2(\vartheta_{02}) \ = \ T_1(\vartheta_{01}) \;\;\;\; \Rightarrow \;\;\;\;  T_2(\vartheta_{2}) \ = \ T_1(\vartheta_{1}) \, .
\end{eqnarray}
The latter demonstrates that the $\{T\}$ scale is an absolute temperature scale, as it is independent of the chosen empirical temperature scale (up to a choice of units). 
Therefore, it remains unaffected by the material properties of the thermometer or the composition of the heat bath.

Combining the preceding discussion with Eqs.~\eqref{eq:46} and~\eqref{eq:48}, the first law of thermodynamics for the entire class of group entropies can be expressed as:
%
%
\beq \label{FLT}
dU  \ = \ \frac{W'(S_{\alpha})}{W(S_{\alpha})} \ \! \frac{T_{\alpha}}{\eta } \ \!dS_{\alpha}  \ + \ \dbar  W
\, , 
\eeq
where $\dbar  W =  -  p_{\rm{phys}} \ \!dV + \cdots$ corresponds to the work done on the system.



\section{The stretched exponential entropy and the black-hole thermodynamics \label{VII.cv}}
As discussed in Sec.~\ref{Intro}, the stretched-exponential entropies form a notable subclass of group entropies that describe thermodynamic systems with sub-exponential state-space growth. As further discussed in Sec.~\ref{Sec.II}, such scaling often arises in systems with long-range interactions or strong correlations, in both classical and quantum settings. 

There are several indications that black-hole thermodynamics may be naturally described within the framework of stretched-exponential entropy~(\ref{3.kk}). A first indication follows from the black-hole area law, according to which the entropy is proportional to the horizon area. Assuming Boltzmann's relation $S=\log W$, the area law implies a holographic state-counting behaviour of the form $W\propto \exp(L^{2})=\exp(V^{2/3})$. Here $V$ is to be interpreted as the thermodynamic volume rather than the geometric interior volume of the black hole~\cite{Note1}. 
Under this identification, the area-law scaling suggests an effective stretched-exponential growth $W(N_{\rm eff})\sim \exp(N_{\rm eff}^{2/3})$, provided one assumes that the relevant effective degrees of freedom scale extensively with $V$. The precise status of this argument is, however, not entirely clear, as it relies on Boltzmann's entropy formula, whose validity is strictly guaranteed only for exponentially growing state spaces.

Secondly, it has been argued that for black holes the asymptotic growth of the number of microstates at fixed energy $E$ follows a stretched-exponential form, $\Omega(E)\sim \exp(E^{\gamma_{\rm{E}}})$, see, e.g.,~\cite{Majumdar_2025}. This behaviour is consistent with 
{various extended entropy relations,
such as those given by Barrow's black-hole entropy~\cite{Capozziello2025,Barrow}, where it is  assumed that the entropy of static black holes scales as a power of the Schwarzschild horizon area $A$. Namely, cf. Eq.~(\ref{barrow})
\begin{equation}
    S_{{\mathrm{B}}} \ = \ \Big(\frac{A}{4l^2_{\rm P}}\Big)^{1+\frac{\Delta}{2}}\, ,
\end{equation}
where $l_{\rm P}$ is the Planck length. The area  
\begin{eqnarray} 
A\ = \ 4\pi r_{\rm S}^2\, ,
\end{eqnarray}
is given in terms of the Schwarzschild radius 
\begin{equation}
    r_{\rm S} \ = \ \frac{2G_{\rm N}M}{c^2}\, ,
\end{equation}
where $G_{\rm N}$ denotes Newton's gravitational constant, $c$ is the speed of light, and $M$ is the mass of the black hole. At this stage, we can express the entropy in terms of the energy by using Einstein's relation $E=Mc^2$ and obtain
\begin{equation}
    S_{{\mathrm{B}}}  \ = \ \eta _{\Delta} E^{2 + \Delta}\; \;\; {\rm where}\;\;\;  \eta _{\Delta} \ = \  \left(\frac{4\pi G_{\rm N}^2}{l^2_{\rm P}c^8}\right)^{1+\Delta/2}\, .
    \label{bh_ent-eng}
\end{equation}
In Sec.~\ref{Sec.II}, we used the extensivity condition~(\ref{ext_cond}) to determine the function $\mathcal{G}(t)$. Alternatively, one can determine the entropy in Eq.~(\ref{eq:Z}) from Eq.~(\ref{bh_ent-eng}) by insisting that in the microcanonical ensemble $p_i=1/\Omega(E)$, in which case we get 
\begin{equation}
    S_{\mathcal{G},\alpha}[1/\Omega(E)] \ = \  \frac{\mathcal{G}(\log(\Omega(E)^{1-\alpha}))}{1-\alpha} \ \propto \ E^{\zeta_{\rm{E}}}\, ,
\end{equation} 
where $\zeta_{\rm{E}}=2+\Delta$ according to Eq.~(\ref{bh_ent-eng}). 
We can now determine the functional form of $\mathcal{G}(t)$ in terms of the inverse $\Omega^{-1}(t)$ of the number of microstates $\Omega(E)$. It can be checked that this leads to the same functional expression for the entropy as in Eq.~(\ref{3.kk}) 
with $\Omega(E)$ substituted for $W(N)$ and $\gamma\equiv \gamma_{\rm{E}}/\zeta_{\rm{E}}$.

For completeness, we stress that we are not aware of any first-principles derivation of the scaling exponent $\gamma$. 
Nevertheless, as argued above, although neither $W(N)$ nor $\Omega(E)$ is presently known with certainty from the fundamental theory of black holes, the entropic form given in Eq.~(\ref{3.kk}) may be regarded as a compelling candidate for black-hole entropy. This is indeed supported by recent results which suggest that $\gamma= 1/2$, reproduces in the whole probability space exactly the same non-additive composition rule for the merging of black-holes recently derived in \cite{Volovik2026} for the microcanonical ensemble. We emphasise that the group entropy in Eq. (\ref{3.kk}) possesses  a robust axiomatic foundation and is fully consistent with the laws of thermodynamics. 
This stands in contrast to several recent approaches to black-hole thermodynamics based, for example, on R{\'e}nyi or Tsallis  entropies, see, e.g.,~\cite{Sadeghi2026,Ahmed2025,Nakarachinda2025,Bhattacharjee2025}, which are generally non-extensive for non-exponential  $W(N)$, and therefore do not naturally reproduce standard thermodynamic consistency conditions.

\subsection{Microcanonical heat capacity}

We now consider two examples illustrating the role of stretched-exponential entropies in black-hole physics. To this end, we rewrite the first law of thermodynamics~(\ref{FLT}) for the case of the stretched-exponential entropy $S_{\alpha,\gamma}$. In this case,~(\ref{FLT}) is tantamount to 
\begin{eqnarray}
dU &=& \ \frac{T_{\alpha,\gamma}}{\eta  } \gamma \left(S_{\alpha,\gamma}\right)^{\gamma -1}  \ \!dS_{\alpha,\gamma} \ + \ \dbar  W \, . 
\label{V.50.kl}
\end{eqnarray}
Let us focus on the heat capacity associated with the stretched-exponential scaling of $W$. We will perform the computation in the microcanonical ensemble. The microcanonical version of~(\ref{V.50.kl}) is given by Eq.~\eqref{VI.87.kl}, and in the present case, it reads:
\begin{eqnarray}
\frac{T_{\alpha,\gamma}}{\eta  } \gamma \left(S_{\alpha,\gamma}\right)^{\gamma -1}  \ \!dS_{\alpha,\gamma} \ = \ dE \ + \ \sum_\nu \bar{X}_\nu d\bar{x}_\nu\, .
\end{eqnarray}
Note that, in the microcanonical ensemble, $S_{\alpha,\gamma}$ is independent of $\alpha$; therefore, the notation $S_{\gamma}$ is sufficient. Moreover, since
\begin{eqnarray}
\frac{\eta  }{\gamma} \frac{S_{\gamma}^{\,1-\gamma}}{T_{\gamma}}
    \ = \ \left( \frac{\partial S_{\gamma}}{\partial E} \right)_{\bar{x}_{\rm ap}}\, ,
    \label{B3.kk}
\end{eqnarray}
(here the subscript $\bar{x}_{\rm ap}$ indicates a fixed value of the applied external parameters),  the absolute temperature depends only on the parameter $\gamma$.

The heat capacity is simply
\begin{eqnarray}
C_{\bar{x}_{\rm ap}} \ = \ \left(\frac{\partial E}{\partial T_{\gamma}}\right)_{\bar{x}_{\rm ap}}\, .
\label{heatcap2}
\end{eqnarray}
To compute this, we differentiate Eq.~(\ref{B3.kk}) once more with respect to $E$. This gives
\begin{eqnarray}
\left( \frac{\partial^2 S_{\gamma}}{\partial E^2} \right)_{\bar{x}_{\rm ap}} &=& \frac{\eta  }{\gamma} \left[ - \frac{S^{1-\gamma}_{\gamma}}{T^2_\gamma} \left( \frac{\partial T_{\gamma}}{\partial E} \right)_{\bar{x}_{\rm ap}} \right. \nonumber \\[2mm] && \left. \ +  \ (1-\gamma ) \ \!\frac{S^{-\gamma}_{\gamma}}{T_\gamma} \left( \frac{\partial S_{\gamma}}{\partial E} \right)_{\bar{x}_{\rm ap}} \right] ,
\end{eqnarray}
which implies that
\begin{eqnarray}
&&\mbox{\hspace{-4mm}}\left( \frac{\partial T_{\gamma}}{\partial E} \right)_{\bar{x}_{\rm ap}} \ = \ C^{-1}_{\bar{x}_{\rm ap}} \nonumber = -\frac{\gamma}{\eta  }\frac{T_{\gamma}^2}{S^{1-\gamma}_{\gamma}}\Bigg[\left( \frac{\partial^2 S_{\gamma}}{\partial E^2} \right)_{\bar{x}_{\rm ap}}  \!\! - \ \\[2mm]
&& \frac{\eta  }{\gamma}(1-\gamma ) \ \!\frac{S^{-\gamma}_{\gamma}}{T_\gamma} \left( \frac{\partial S_{\gamma}}{\partial E} \right)_{\bar{x}_{\rm ap}}\Bigg] \nonumber \\[2mm]
&&= \frac{\eta  }{\gamma} S^{-\gamma} \left[1-\gamma -\frac{SS''}{(S')^2} \ \right], \label{heat_capa}
\end{eqnarray}
where we used the short-hand notation
\begin{eqnarray}
S' \ \equiv \  \left( \frac{\partial S_{\gamma}}{\partial E} \right)_{\bar{x}_{\rm ap}}\, ,
\end{eqnarray}
and similarly for $S''$.

Let us remark on the applicability of this expression. When we derived the equilibrium condition and introduced the temperature $T_{\alpha,\gamma}$ in Sec. \ref{Section VI} and \ref{Sec.III.aaa}, we worked under the assumption of energy being additive and entropy being non-additive. This raises the question if Eq. (\ref{heat_capa}) can be of relevance to black holes involving non-additive  gravitational energy. We propose that Eq.~(\ref{heat_capa}) can be valid even if the equilibrium condition derived in Sec. \ref{Sec.III.aaa} does not apply to black holes. This is because  to establish the temperature in Eqs.~(\ref{B3.kk}) and (\ref{heatcap2}) we just need to bring the black hole in contact with a weakly interacting thermometer, for which the combined system of black hole and thermometer is described by the Cartesian product of the state spaces of the two systems. In our case this implies a device described by the stretched exponential group entropy but only interacting sufficiently strongly with the black hole to establish equilibrium but not doing this via long-range gravitational forces.  

To understand whether the heat capacity is positive or negative, we need to know the explicit form of $S_\gamma$ as a function of $E$.

From Eq.~(\ref{I.10.kl}) we may express the entropy in the form $S_{\gamma}=N(E)$, where $N(E)$ denotes the resulting energy distribution. 
This representation enables us to analyse the conditions on $N(E)$ under which the heat capacity~\eqref{heat_capa} becomes negative. 
We assume that $N(E)>0$. 
Eq.~\eqref{heat_capa} then implies that the condition $C_{\bar{x}_{\rm ap}}<0$ is equivalent to
\begin{eqnarray}
N(E) N''(E) \ - \ (1-\gamma) [N'(E)]^2 \ > \ 0\, .
\end{eqnarray}
To analyse this inequality, we introduce the auxiliary function $h(E)\equiv N'(E)[N(E)]^{\gamma-1}$. With this we have 
\begin{eqnarray}
&&\mbox{\hspace{-4mm}}h'(E) \nonumber \\[2mm]  
&&\mbox{\hspace{-4mm}} = \  N(E)^{\gamma-2} [N(E) N''(E) \ - \ (1-\gamma) (N'(E))^2 ]\, .~~~~~~
\end{eqnarray}
Since $N(E)^{\gamma-2}>0$, we deduce that 

\begin{eqnarray}
C_{\bar{x}_{\rm ap}} \ < \ 0 \;  \Leftrightarrow \; \frac{d}{dE}[N'(E) N(E)^{\gamma-1}]\ > \ 0\, ,
\end{eqnarray}
which can be rewritten, equivalently, as
\begin{eqnarray}
C_{\bar{x}_{\rm ap}} \ < \ 0 \;  \Leftrightarrow \;  \frac{1}{\gamma} [N(E)^{\gamma}]''\ > \ 0\, .
\label{VII.106.ck}
\end{eqnarray}
This inequality constrains the class of energy distributions $N(E)$ that are compatible with a negative heat capacity. 
For example, for quasi-additive energies one may assume a polynomial scaling $N(E)\propto E^{c}$, in which case 
Eq.~(\ref{VII.106.ck}) implies the condition $c\,\gamma>1$. In particular, by employing our former argument, we have
\begin{eqnarray}
c \ = \  \gamma_{\rm{E}}/\gamma \ = \  \zeta_{\rm{E}} \ = \  2 +  \Delta\, .
\end{eqnarray}
In addition, for $\gamma = 2/3$, one finds that $c\gamma > 1$ for all admissible $\Delta$, implying that the heat capacity is negative. While this behaviour is a characteristic feature of black holes, it should be noted that, in the present case, the entropy remains extensive.  Note that $C<0$ implies that the stationary point of $S_{\alpha}^{\rm Tot}$ found in Sec.~\ref{Section VI} is not a maximum, consistently with the well-known microcanonical instability of self-gravitating systems~\cite{Rufo:14,LB,Thirring}.

In passing, we emphasize that $N_{\rm eff}$, introduced at the beginning of this section, should not be interpreted as a microscopic particle number. Rather, it represents an effective extensive parameter characterizing the asymptotic growth of the black-hole state space. For the Schwarzschild black hole considered here, we identify $N_{\rm eff}\propto V_{\rm th}$, where $V_{\rm th}$ is the thermodynamic volume, and this identification leads to the stretched-exponential scaling $W(N_{\rm eff})\sim \exp(N_{\rm eff}^{2/3})$. Furthermore, since the corresponding entropy $S_{\alpha,2/3}$ scales as $V_{\rm th}/V_{\rm P}$  ($V_{\rm P}$ is a reference Planck volume), the quantity $N_{\rm eff}$ may be regarded as an effective count  of microscopic information cells contained within the thermodynamic volume --- one bit of information is associated with a volume of order one Planck volume. This interpretation is reminiscent of Bekenstein's original argument that the number of fundamental microstates associated with a black hole should scale with its horizon area, with the precise proportionality constant determined subsequently by Hawking's calculation of black-hole radiation.

\subsection{Black-body radiation in a black-hole spacetime\label{VII.B.kk}}
%
As a second example, we consider the thermal radiation of an electromagnetic (EM) field inside a black-body cavity placed in the gravitational field of a static black hole. Operationally, the EM black-body radiation can be characterized by measuring the EM spectrum in a cavity at a fixed proper distance from the horizon, as locally measured by a static observer co-moving with the cavity. In the following, we focus on two specific cavity locations: one in the near-horizon region and one far from the horizon.

We assume that the cavity has perfectly reflecting walls, so that it behaves as an idealized black body with a stationary thermal radiation spectrum determined solely by the cavity temperature and independent of any microscopic details of the walls. In addition, we assume that the cavity is sufficiently small compared to curvature scales, so that tidal effects can be neglected.
Let the cavity have radial thickness $\Delta \rho$ and transverse area $A$, so that the proper volume is $V_{\rm prop} = A \Delta \rho$. In Appendix~\ref{ap.b}, we show that the microcanonical density of EM field states with Killing energy below $E_{\infty}$ (i.e., energy measured at infinity) is

\begin{widetext}
\begin{eqnarray}
\Omega_{\rm{TC}}(E_{\infty},\rho_0, \Delta \rho) \ \sim \ \left\{
  \begin{array}{ll}
    \exp\left[\frac{4}{3} A \left( \frac{\pi^{2} }{30 \kappa^{3} \rho_{0}^{2}} \right)^{1/4} \!\sigma_{\infty}^{3/4}\right] , & \;\;\;\;\; \rho_0 \ll \Delta \rho \;\;\;\;\;\hbox{(near~horizon)}\, ; \\[3mm]
    \exp\left[\frac{4}{3}V_{\rm prop}\left(\frac{\pi^2}{15}\right)^{1/4} \!u_{\infty}^{3/4}\right], & \;\;\;\;\; \rho_0  \gg \Delta \rho \;\;\;\;\; \hbox{(far from~horizon)}\, ,
  \end{array}
\right.
\label{VII.108.kk}
\end{eqnarray}
\end{widetext}

where the subscript ``$\rm{TC}$'' denotes ``total cumulative''.
Here, $\rho_0$ is the proper distance of the cavity from the horizon,
$\kappa$ is the surface gravity of the black hole, and
$\sigma_{\infty}$ and $u_{\infty}$ denote the energy per unit transverse
area and the energy density of the box as measured at infinity,
respectively.

Equation~(\ref{VII.108.kk}) states that the density of states is strongly weighted toward configurations localized near the horizon, such that the dominant contribution to the entropy scales as
$\exp(\alpha A)$ rather than $\exp(\alpha V_{\rm{prop}})$ when the cavity is sufficiently close
to the horizon.

Since we have shown that the entropies $S_{\alpha,\gamma}$ constitute a fully consistent set of thermodynamic entropies, and since they belong to the Uffink class, one can derive within the MaxEnt framework a statistical ensemble equivalent to the Gibbs canonical ensemble. Imposing the constraints of probability normalization and fixed average energy, the MaxEnt distribution is obtained by extremizing $S_{\alpha,\gamma}$ subject to these constraints. 
Because $S_{\alpha,\gamma}$ is a positive monotonic function of the R\'enyi entropy, the resulting MaxEnt distribution under standard linear constraints is identical to the R\'enyi MaxEnt distribution, which is known to be (see ~\cite{JKZ} and also Appendix~\ref{Appendix-C}):

%
\begin{eqnarray}
p_k \ = \ \frac{1}{Z} \left\{1 \ + \ (1-\alpha) \beta^{\rm{R}}(U) [E_k - U] \right\}^{1/(\alpha -1)} \, ,~~~
\label{B.110.df}
\end{eqnarray}
with $Z$ being the ensuing partition function and $U$ the corresponding average value of energy. The parameter $\beta^{\mathrm{R}}$ plays the role of the R\'enyi ``inverse
temperature''. It is related to the Lagrange multiplier $b$ associated with the
average-energy constraint in the MaxEnt formulation through the relation
$\beta^{\mathrm{R}} = b/ \alpha$, see~\cite{J-Temp,JKZ}  and also Appendix~\ref{Appendix-C}. An analogous relation holds for systems with a continuous energy spectrum.

As reviewed in \cite{Beck2009}, monotonic transformations of the R\'enyi entropy necessarily preserve its extremizing distributions under identical constraints. Consequently, within the present thermodynamic framework, the canonical distributions are given by $q$-exponentials (power laws). It is worth noting that, despite the stretched-exponential phase-space scaling, $W(N)\!\sim\!\exp(N^\gamma)$, characterizing the microcanonical regime, this framework does not automatically generate more complex canonical profiles, such as stretched $q$-exponentials, which are often relevant in phenomena like turbulence. Obviously, the values of the entropy computed on the $q$-exponential distribution will be different for different group entropies.

It is now convenient to fix the temperature units by choosing $\eta=\gamma$;
see Eqs.~(\ref{V.50.kl}) and (\ref{T=theta}). With this choice, the inverse absolute
temperature satisfies relation
%
%
%
\begin{eqnarray}
\beta^*_{\alpha,\gamma} \ = \  \left(S_{\alpha,\gamma}\right)^{\gamma -1} \left(\frac{\partial S_{\alpha,\gamma}}{\partial U}  \right)_V \ = \ {{\frac{1}{\gamma}}} \left(\frac{\partial S^{\rm{R}}_{\alpha}}{\partial U}  \right)_V\, ,~~~
\label{VII.b.11.bh}
\end{eqnarray}
where $S^{\rm{R}}_{\alpha}$ is R\'enyi entropy of order $\alpha$. So, for the on-shell (i.e., thermodynamic) entropy and internal energy we have
\begin{eqnarray}
\beta^*_{\alpha,\gamma} \ &=& \ {{\frac{1}{\gamma}}} {\left(\frac{\partial S_{\alpha}^{\rm{R}}}{\partial \beta^{\rm{R}}}  \right)_V}\Big/{\left(\frac{\partial U}{\partial \beta^{\rm{R}}}  \right)_V} \nonumber \\[2mm] &=& \ \frac{b}{{\gamma}} \ = \ \frac{\alpha}{{\gamma}} \ \! \beta^{\rm{R}}(U).~~~~
\end{eqnarray}
One can equivalently work with the shifted distribution
\begin{eqnarray}
p_k \ = \ \frac{1}{Z} \left[1 \ + \ (1-\alpha) \beta^{\rm{R}}(0) E_k \right]^{1/(\alpha -1)} \, ,
\label{B.114.ck}
\end{eqnarray}
in which case the parameter  $\beta^{\rm{R}}(0)$ is related to the parameter $\beta^{\rm{R}}(U)$ and the absolute temperature via the relations~\cite{JKZ}
\begin{eqnarray}
\beta^{\rm{R}}(0) \ &=& \ \frac{\beta^{\rm{R}}(U)}{1 \ + \ \beta^{\rm{R}}(U) (\alpha - 1) U} \nonumber \\[2mm]
&=& \ \frac{\gamma}{\alpha}\frac{\beta^{\rm{*}}_{\alpha,\gamma}}{1 \ + \ \gamma\ \! \beta^{\rm{*}}_{\alpha,\gamma} (1- 1/\alpha) U} \, .
\label{B.hj}
\end{eqnarray}
We now define the inverse physical temperature and the energy measured at $r\rightarrow \infty$ as
$\beta^{*}_{\alpha,\gamma}(U_{\infty})$ and $U_{\infty}$, respectively.
Analogously, we denote the corresponding quantities evaluated at the vicinity of the horizon (parametrized by the radial coordinate) as
$\beta^{*}_{\alpha,\gamma}(U_{\mathrm{h}})$ and $U_{\mathrm{h}}$, respectively. 
The energies $U_{\mathrm{h}}$ and $U_{\infty}$ are related by the gravitational redshift relation~\cite{Wald:84}
\begin{eqnarray}
U_{\mathrm{h}}(r) \ = \ \frac{U_\infty}{\sqrt{-g_{00}(r)}}\, ,
\label{B.115.kl}
\end{eqnarray}
which holds mode by mode and therefore also for expectation values (provided the cavity in question is sufficiently small).  In writing~(\ref{B.115.kl}), we used the $(-,+,+,+)$ metric signature. 
In particular, for the Schwarzschild black-hole spacetime, the time-time component of the metric tensor is given by
\begin{eqnarray}
g_{00}(r) \ = \ -\left(1 -  \frac{2 G_{\rm N} M}{r}\right) \, .
\end{eqnarray}
Consequently, the gravitational redshift formula~(\ref{B.115.kl}) states that the measured energy is higher the deeper one is within the black-hole gravitational potential.

From~(\ref{VII.b.11.bh}), we might thus write
\begin{eqnarray}
\beta^*_{\alpha,\ \! 2/3}(U_{\rm{h}}) \ &=& \ \frac{3}{2} \left(\frac{\partial S^{\rm{R}}_{\alpha}}{\partial U_{\rm{h}}}  \right)_V \nonumber \\[2mm] &=& \   \frac{3}{2} \sqrt{-g_{00}(r) }\ \!  \left(\frac{\partial S^{\rm{R}}_{\alpha}}{\partial U_{\infty}}  \right)_V \nonumber \\[2mm] &=& \ \frac{3}{2} \sqrt{-g_{00}(r) }\ \! \beta^*_{\alpha,\ \! 1}(U_{\infty})\, .
\label{B.117.gh}
\end{eqnarray}
In this derivation, we have followed~(\ref{VII.108.kk}) and assumed that $\gamma = 1$ for the $r \rightarrow \infty$  region and $\gamma = 2/3$ in the region close to the horizon. The reader can readily generalize these results to any preferred choice of $\eta$, corresponding to changing the temperature scale.

Relation~(\ref{B.117.gh}) can be equivalently rewritten as
\begin{eqnarray}
T_{\alpha, \ \! 2/3}(U_{\rm{h}}) \ = \ \frac{2}{3} \frac{T_{\alpha, 1}(U_{\infty})}{\sqrt{-g_{00}(r) }}\, .
\label{B.118.ch}
\end{eqnarray}
%
%
%
%
Note that the factor $\gamma$ reflects the choice $\eta=\gamma$ made above. Equation~(\ref{B.118.ch}) generalizes the Tolman--Ehrenfest relation~\cite{tolman} for the temperature transformation in a static gravitational field in terms of the absolute temperatures naturally associated with the two scaling regimes.

We note in passing that the area-law scaling in Eq.~(\ref{VII.108.kk}) has the same underlying origin as in 't~Hooft's brick-wall model for black holes~\cite{tHooft1,tHooft2}. In the latter approach, the area law arises from the accumulation of quantum field modes in the near-horizon region, with the Bekenstein--Hawking entropy recovered once short-distance divergences are regulated by Planck-scale physics. In the present framework, the area law likewise originates from the enhancement of degrees of freedom close to the horizon. However, no fundamental ultraviolet regulator is introduced; the only assumption is that $\rho_0 \ll \Delta\rho$ and that $\rho_0 >0 $. The precise value of $\rho_0$, including whether it is associated with the Planck scale $l_{\rm{P}}$ or not, is therefore not essential for our argument.

\subsection{Stefan--Boltzmann law}

As a continuation of the preceding analysis, we now derive an analogue of the Stefan--Boltzmann law describing the thermal radiation of an EM field emitted by a black body in a static black-hole spacetime. This analysis will provide further insight into the physical interpretation of the absolute temperature $T_{\alpha,\gamma}$.

We begin by recalling that a thermodynamically consistent entropy can be obtained from $S_{\alpha,\gamma}$ by applying the MaxEnt prescription. The latter relies on two constraints: normalization of the probability distribution and the specification of the mean energy density. In particular, when the cavity is located far from the horizon, the averaged energy density takes the form 
\begin{eqnarray}
u(T_{\alpha,1}) \ = \  \langle u \rangle_{\alpha,1}  \ = \ 3\ \!p_{\mathrm{phys}}  \, .
\label{VII.82.klm}
\end{eqnarray}
Here, the average $\langle\ldots \rangle_{\alpha,1} $ is computed with respect to the MaxEnt distribution~(\ref{B.110.df}). We recall here that  $p_{\mathrm{phys}}$ also depends on $\alpha$ (and, in general, also on $\gamma$) as stressed in Sec.~\ref{Sec.III.aaa}.
Constraint~(\ref{VII.82.klm}) arises from the fact that the trace of the energy-momentum tensor for the electromagnetic field vanishes --- a consequence of the global scale invariance of the electromagnetic action --- and from the assumption that the radiation is contained in an isotropic environment (cavity), in which case the spatial components of the energy-momentum tensor are identical (factor 3) and can be identified with the physical pressure.

We further assume that in thermal equilibrium the internal energy $U$ of the radiation is distributed uniformly throughout the cavity, meaning that
\begin{eqnarray}
U(T_{\alpha,1},V ) \ \equiv \ \langle U \rangle_{\alpha, 1} \ = \ V \langle u \rangle_{\alpha,1} \ = \  V u(T_{\alpha,1})\, .~~~
\label{VII.82.kl}
\end{eqnarray}
We now return to Eq.~(\ref{IV.40.cc}) and substitute the expressions for 
$p_{\mathrm{phys}}$ and $U$. A straightforward calculation yields
\begin{eqnarray}
T_{\alpha,1}\,\frac{d u(T_{\alpha,1})}{d T_{\alpha,1}}
  \ = \  4\,u(T_{\alpha,1})\, ,
\end{eqnarray}
whose solution is
\begin{eqnarray}
u(T_{\alpha,1}) \ =\  \sigma_{\alpha,1}\, T_{\alpha,1}^4\, ,
\label{B4.klc}
\end{eqnarray}
where $\sigma_{\alpha,1}$ is an integration constant.

{
In Appendix~\ref{appendix:D}, we show that for the density of energies $u_{\infty}$ and $u_{\rm{loc}}$, 
one has that
\begin{eqnarray}
\frac{u_{\rm{loc}}}{u_{\infty}} \ = \ (-g_{00})^{-2}\, .
\label{124.hk}
\end{eqnarray}
By coupling this with the Tolman--Ehrenfest 
relation~(\ref{B.118.ch}), we obtain that
\begin{eqnarray}
u_h  \ &\equiv& \ u_{\rm{loc}}(r_h) \ = \ u(T_{\alpha, 2/3}) \nonumber \\[2mm] &=& \  \sigma_{\alpha,1} \left(\frac{3}{2}\right)^{\!4} T^4_{\alpha,2/3} \, .
\label{124.kl}
\end{eqnarray}
Results~(\ref{B4.klc}) and~(\ref{124.kl}) are nothing but generalized versions of the Stefan--Boltzmann law. 
In particular, Eq.~(\ref{124.kl}) implies that the Stefan--Boltzmann constants in the near-horizon region and in the asymptotic far-field region are related through the expression
\begin{eqnarray}
\sigma_{\alpha,2/3}  \ = \  \sigma_{\alpha,1} \left(\frac{3}{2}\right)^{\!4}\, .
\label{C.125.kl}
\end{eqnarray}
This can again be easily generalized to any choice of the scale factor $\eta$.
In passing, we note that Eq.~(\ref{C.125.kl}) implies that the constant $\sigma_{\alpha,2/3}$ is independent of the particular choice of the short-distance parameter $\rho_0$, since $\sigma_{\alpha,1}$, being
defined in the asymptotic region, does not depend on $\rho_{0}$ either. The apparent dependence on $\rho_0$ entering through $g_{00}$ is exactly cancelled once the generalized Tolman--Ehrenfest relation is taken into account. This does not imply, however, that the generalized Stefan--Boltzmann law Eq.~(\ref{124.kl}) is independent of $\rho_0$. On the contrary, $\rho_0$ enters Eq.~(\ref{124.kl}) implicitly through  $T_{\alpha,2/3}$, whose value depends on $\rho_0$ via the generalized Tolman--Ehrenfest relation.

Eq.~(\ref{C.125.kl}) merely establishes a relationship between generalized Stefan--Boltzmann constants that correspond to different state-space scaling exponents $\gamma$. Provided that the scaling regimes $\gamma = 1$ and $\gamma = 2/3$ are realized, this relation is independent of the short-distance parameter $\rho_0$. This situation is somewhat reminiscent of the notion of universality in the renormalization group, where at a renormalization-group fixed point (in our case, when the system reaches a given $\gamma$ scaling regime), the scaling behaviour of various physical observables becomes insensitive to the microscopic details of the underlying theory, depending only on the universality class.

By making use of Eq.~(\ref{29ccg}), one may express the physical pressure as
%
%
\begin{eqnarray}
p_{\mathrm{phys}}
\ = \ \frac{T_{\alpha,\gamma}}{\eta } \gamma \left(S_{\alpha,\gamma}\right)^{\gamma-1}
\left(\frac{\partial S_{\alpha,\gamma}}{\partial V}\right)_{\!U}\, .
\label{C.126.hj}
\end{eqnarray}
This relation allows one to determine the temperature $T_{\alpha,\gamma}$ once the physical pressure $p_{\mathrm{phys}}$ is known. 
Because the parameter $\alpha$ 
is associated with the R\'{e}nyi-entropy part of  $S_{\alpha,\gamma}$, it
quantifies correlations within the system~\cite{J-K:19,Jizba-Dun:16},  whereas $\gamma$ characterizes the scaling of the state space. 
Accordingly,  the physical pressure appearing in Eq.~(\ref{C.126.hj}) may be interpreted as a \emph{correlation pressure} --- for instance, the excess or virial (configurational) pressure~\cite{McQuarrie} --- that is sensitive only to a specified class of correlations. When $p_{\mathrm{phys}}$ corresponds to the total pressure with all correlations fully accounted for,  one recovers $\alpha = 1$.
Note that $\sigma_{1,1}$ coincides with the conventional Stefan--Boltzmann constant.


\subsection{Some remarks}

Let us close with a few remarks. 
First, within the canonical formulation the absolute temperature exhibits a dependence on both the parameters $\alpha$ and $\gamma$.
This stands in clear contrast to the microcanonical ensemble, where the temperature depends only on $\gamma$ [see Eq.~(\ref{B3.kk})]. This reflects the inequivalence between ensembles (known to be present for systems with long-range interactions) and points towards the fact that energy fluctuations cannot be neglected in the canonical ensemble. The parameter $\alpha$ may therefore be interpreted as encoding additional information about the system that is not accessible at the microcanonical level, namely correlations~\cite{J-K:19}.

Second, we can rewrite Eq.~(\ref{V.50.kl}) equivalently in the form
\begin{eqnarray}
dU &=& \   \frac{T_{\alpha,\gamma}}{\eta }  \ \! d\left[\left(S_{\alpha,\gamma}\right)^{\gamma}\right]\ + \ \dbar  W \, .
\label{V.56.klm}
\end{eqnarray}
By setting $\eta = 1$ to fix the temperature scale, we recover a relation analogous to the first law of thermodynamics, albeit expressed in terms of a non-extensive ``entropy'' $(S_{\alpha,\gamma})^{\gamma}$}.

We now  consider a state space that exhibits stretched-exponential growth  of the form $W(N)\propto \exp(L^{2})=\exp(V^{2/3})$, i.e., holographic scaling. To obtain an extensive entropy --- necessary for a consistent thermodynamic formulation --- the stretched-exponential entropy must be specified with the parameter choice $\gamma = 2/3$. In this case, however, $(S_{\alpha,2/3})^{2/3}$ appearing in Eq.~(\ref{V.56.klm}) scales as $L^{2}$. 
Consequently, Eq.~(\ref{V.56.klm}) recovers the structure of the ``first law'' as it appears in black-hole thermodynamics~\cite{Bekenstein,Hawking,Hawking2}, even though $(S_{\alpha,\gamma})^{\gamma}$ cannot constitute a genuine thermodynamic entropy because it is not extensive and it does not belong to the class of group entropies.

Third, from a thermodynamic axiomatic standpoint (in the sense of Carath\'{e}odory or Clausius), one is not immediately allowed to equate the Hawking (formal) temperature with the absolute temperature, because the foundational assumptions (extensivity, additivity, Carnot cycle construction) fail. Nevertheless, the present construction demonstrates that the temperature $T_{\alpha,\gamma}$ can indeed be regarded as a genuine absolute temperature. Since, for the area-law scaling of the state space,  Eq.~(\ref{V.50.kl}) corresponds to an analog of the first-law of black-hole thermodynamics, we see 
that the Hawking temperature may be consistently interpreted as an absolute  temperature --- namely $T_{\alpha,\gamma}$, even though it differs from the conventional Clausius--Kelvin absolute temperature.

Fourth, since  $(S_{\alpha,\gamma})^{\gamma}$ is non-extensive  (in case of sub- or super-exponential state-space scaling), the corresponding absolute temperature $T_{\alpha, \gamma}$ is not intensive.
This is consistent with the result known within more traditional thermodynamic approaches to systems with long-range interactions~\cite{Rufo:14}, and is also consistent with conclusions drawn in a variety of cosmological contexts based on the holographic (or generalized holographic) principle~\cite{Padmanabhan:90}.
\section{Conclusions \label{conclusions}}


In this work, we have formulated a thermodynamic framework for the group entropies introduced recently in Refs.~\cite{PT2011PRE,T2016AOP,Tempesta2016}, treating them as extensive functionals defined on probability spaces of microstates. These entropies can be classified into distinct universality classes, characterized by the asymptotic scaling behaviour of the number of accessible microstates $W(N)$, with $N$ representing the number of constituents  (or the number of available degrees of freedom). Whenever the growth of $W(N)$ departs from a purely exponential form, 
one is naturally led to generalized formulations of Boltzmann--Gibbs statistical thermodynamics.
Within this framework, we derived a corresponding zeroth law of thermodynamics and proved that, for all group entropies, an empirical physical temperature can be consistently defined. This, in turn, allowed us to establish the condition for thermal equilibrium between different subsystems within the same class $W(N)$. Furthermore, by employing Carath{\'e}odory's axiomatic formulation of thermodynamics, we established the existence of an absolute thermodynamic temperature for all group entropies. We also showed that the empirical temperature obtained coincides with an absolute temperature, up to a multiplicative constant corresponding to the choice of temperature units. It has also been shown that, for a non-degenerate ground state, group entropies satisfy Falk's formulation of the third law of thermodynamics.

As an application of the thermodynamic framework developed here, we investigated black-hole thermodynamics within the class of group entropies associated with the stretched-exponential scaling of $W(N)$. In recent years, generalized entropies have become an active area of research in black-hole thermodynamics, where they have been explored as possible extensions of the Bekenstein--Hawking entropy motivated by quantum-gravitational effects, non-trivial correlations, and modified microscopic state counting~\cite{PhysRevD.111.024013,PhysRevD.104.084030,PhysRevD.102.064014,Bhattacharjee2025,SUCU2026170494}. The present work contributes to this ongoing line of investigation by providing a systematic thermodynamic framework for analysing such generalized entropies.  In particular, we verified that a defining feature of black holes --- namely, their negative specific heat --- is naturally realized within this framework, even though the entropy remains extensive.  This result holds for the stretched-exponential entropies corresponding to both the Bekenstein--Hawking and Barrow entropy scalings. We also analysed black-body radiation, and the resulting generalized Stefan--Boltzmann law, in the vicinity of strongly gravitating black holes, where the adiabatic approximation is well justified. We showed that the resulting absolute temperature belongs to a two-parameter family of temperatures. In contrast to the microcanonical description, the temperature acquires an additional dependence on a parameter $\alpha$, which encodes information about the system that resides in higher-order correlations.

The application of our framework to black holes and other systems characterized by long-range interactions or strong correlations indicates that the resulting thermodynamic behaviour can be understood as a direct consequence of the non-exponential scaling of the number of microstates with the number of degrees of freedom. 


By contrast, the exponential scaling underlying Boltzmann--Gibbs statistical thermodynamics corresponds to a state space that factorizes into a Cartesian product of single-particle state spaces. Deviations from this factorized structure arise from strong interactions or correlations and are encoded in the functional dependence of $W(N)$.

A natural direction for future work is to construct simple microscopic models that reproduce the asymptotic growth of the phase space from the underlying dynamics. Existing examples suggest that microscopic composition rules, global (non-local) constraints, or strong collective correlations can substantially modify the combinatorial growth of the accessible state space. For example, the pairing model~\cite{JPPT2018}  exhibits super-exponential growth through dynamically evolving pairing rules, while Hagedorn spectra in hadronic bootstrap models and string theory~\cite{Frautschi1971,GreenSchwarzWitten1987,Polchinski1998} arise from the combinatorics of constrained collective excitations, leading asymptotically to $W(N)\sim \exp(a\sqrt{N})$, with $a>0$ and $N$ the total oscillator number. Likewise, ordinal-pattern state spaces~\cite{BandtPompe2002} naturally exhibit the super-exponential scaling $W(N)=N!\sim \exp(N\ln N)$. Constructing an analogous exactly solvable microscopic model in which genuine long-range interactions, such as gravitation, directly produce the stretched-exponential scaling $W(N)\sim \exp(N^{\gamma})$, with $0<\gamma<1$, remains an outstanding open problem and an important direction for future work.


\begin{acknowledgments}
P.J. acknowledges support from the Czech Science Foundation Grant (GA\v{C}R), Grant No. 25-18105S.  P.T. was supported by the Project PID2024-156610NB-I00 of Ministerio de Ciencia, Innovaci\'on y Universidades, Spain,  and by  the Severo Ochoa Programme for Centres of Excellence in R\&D (CEX-2023-001347-S), Ministerio de Ciencia, Innovaci\'{o}n y Universidades y Agencia Estatal de Investigaci\'on, Spain. P.T. also thanks the Gruppo Nazionale di Fisica Matematica (GNFM) of the Istituto Nazionale di Alta Matematica (INdAM).
\end{acknowledgments}

\appendix

\section{Formal group laws \label{sec:22}}


In this appendix, we briefly recall some basic facts concerning formal group theory that are required in the main text. For a comprehensive discussion of this topic, the reader is referred to the classic monograph~\cite{Haze}.

We first recall that all rings considered in this paper are associative, commutative unital rings. In particular, given a ring $(R,+,\cdot )$, we
shall denote by $0$ the neutral element of the addition operation $+:R\times
R\rightarrow R$ and by $1$ the neutral element of the multiplication
operation $\cdot :R\times R\rightarrow R$.

\begin{definition} \label{formalgrouplaw}
Let $R$ be a commutative ring with identity, and $R[[
x_{1},x_{2},\ldots ]]$ be the ring of formal power series in the variables 
$x_{1}, x_{2}, \ldots$ with coefficients in $R$. A 
one-dimensional formal group law over $R$ is a formal power series in two
variables $\Phi \left( x,y\right) \in R\ [[  x,y ]]$ of the form $%
\Phi \left( x,y\right) =x+y~+$ \emph{terms of higher degree}, such that 
\begin{eqnarray*}
i)\mbox{\hspace{4mm}} &&\Phi \left( x,0\right) \ = \ \Phi \left( 0,x\right) \ = \ x\, ,\nonumber \\[2mm]
ii)\mbox{\hspace{4mm}} &&\Phi \left( \Phi \left( x,y\right) ,z\right) \ = \ \Phi \left( x,\Phi
\left( y,z\right) \right) \, .
\end{eqnarray*}
When $\Phi \left( x,y\right) =\Phi \left( y,x\right)$, the formal group law
is said to be commutative.
\end{definition}

For any formal group law $\Phi (x,y)$, there exists an \textit{inverse
formal series} $\varphi \left( x\right) $ $\in R[[ x]] $ such
that $\Phi \left( x,\varphi \left( x\right) \right) =0$. This is the reason
why we talk about \textit{formal group laws}. For a thorough
exposition on formal group theory see~\cite{Haze}. Some relevant examples are listed below.
\begin{itemize}
\item The \textit{additive} \textit{group law} 
\begin{equation*}
\Phi (x,y)\ = \ x \ + \ y\, .
\end{equation*}
\item The \textit{multiplicative} \textit{group law} 
\begin{equation*}
\Phi (x,y) \ = \ x \ + \ y \ + \ axy\, .
\end{equation*}
\item The \textit{hyperbolic} \textit{group law} (addition of velocities in
special relativity) 
\begin{equation*}
\Phi (x,y)\ = \ \frac{x\ + \ y}{1\ + \ xy}\, .
\end{equation*}
\item The \textit{Euler group law} 
\begin{equation*}
\Phi (x,y)  =  \Big(x\sqrt{1-y^{4}} +  y\sqrt{1-x^{4}}\Big)/(1  +  x^{2}y^{2})\, .
\end{equation*}
which appears in the sum of elliptic integrals 
\begin{equation*}
\int_{0}^{x}\frac{dt}{\sqrt{1-t^{4}}} \ + \ \int_{0}^{y}\frac{dt}{\sqrt{1-t^{4}}}%
\ = \ \int_{0}^{\Phi (x,y)}\frac{dt}{\sqrt{1-t^{4}}}\, .
\end{equation*}\\[-2mm]
\end{itemize}
%
Let  $B=\mathbb{Z}[b_{1},b_{2},...]$ be the  ring of integral polynomials in infinitely many variables. We shall consider the series
\beq
F\left( s\right) \ = \  \sum_{i=0}^{\infty} b_i \frac{s^{i+1}}{i+1}\, ,
\label{I.1}
\eeq
with $b_0=1$. Let $G\left( t\right)$ be its compositional inverse:
\beq
G\left( t\right) \ = \ \sum_{k=0}^{\infty} a_k \frac{t^{k+1}}{k+1} \, ,\label{I.2}
\eeq
i.e., $F\left( G\left( t\right) \right) =t$. From this property, we deduce $a_{0}=1$, $a_{1}=-b_1$, $a_2= \frac{3}{2} b_1^2 -b_2$, $\ldots$.
The {\em Lazard formal group law} \cite{Haze} is defined by the formal power series
\begin{eqnarray}
\Psi_{\mathcal{L}} \left( s_{1},s_{2}\right) \ = \ G\left( G^{-1}\left(
s_{1}\right) \ + \ G^{-1}\left( s_{2}\right) \right)\, .
\end{eqnarray}
%
The coefficients of the power series 
%
$G\!\left( G^{-1}\!\left( s_{1}\right) \ + \  G^{-1}\!\left( s_{2}\right) \right)$ 
%
lie in the ring $B \otimes \mathbb{Q}$ and generate, over $\mathbb{Z}$, a subring $L \subset B \otimes \mathbb{Q}$, referred to as the \emph{Lazard ring}.

For any commutative one-dimensional formal group law over any ring $R$, there exists a unique homomorphism $L\to R$ under which the Lazard group law is mapped into the given group law (the \textit{universal property} of the Lazard group).

In the construction of a group entropy, the composition law is algebraically
a commutative formal group law, i.e., a  function of the form 
\begin{equation}
\Phi (x,y) \ = \ \sum_{n=0}^{\infty }\sum_{i=0}^{n}a_{i,n-i}x^{i}y^{n-i}\, ,
\label{Phi(x,y)}
\end{equation}%
that satisfies conditions (C2)--(C4) of Definition \ref{Def1}. Therefore, $%
a_{i,n-i}=a_{n-i,i}$ for all $n\geq 1$ and $0\leq i\leq n$ by symmetry, and $%
a_{0,0}=0$ and $a_{n,0}=0$ for all $n\geq 2$ by the null-composability. This
is the origin of the connection between entropic measures and formal group theory \cite{PT2011PRE}.
We stress that for the entropies considered in the main text the composition
law need not be a formal power series in the sense of
Definition~\ref{formalgrouplaw}. For instance, the stretched-exponential
class~\eqref{3.kk} yields $\Phi (x,y) \ = \ \big(x^{\gamma} \ + \ y^{\gamma}\big)^{1/\gamma}$, where $G(t) \propto t^{1/\gamma}$ is not analytic at the origin
for $\gamma<1$. The composition law nevertheless defines a commutative group law on
$\mathbb{R}_{\geq 0}$, which is all that is required for the thermodynamic
construction developed in this work.

 \section{Proof of Eq.~(\ref{VII.108.kk}) \label{ap.b}}

In this appendix, we analyse the dependence of the density of EM field states with the asymptotic (Killing) energy $E_{\infty}$ on the proper distance from the  Schwarzschild black-hole horizon. We distinguish between two regimes: the near-horizon region and the region far from the horizon.

\subsection{Near-horizon region}

Near the horizon, the Schwarzschild metric reduces to the form (Rindler form)
\begin{eqnarray}
ds^2 \ \simeq \ - (\kappa \rho_0)^2 dt^2 \ + \ d\rho^2_0 \ + \ r_h^2 d\Omega^2\, ,
\label{B.1.kh}
\end{eqnarray}
where $d\Omega^2$ is a shorthand for $\left(d\ttheta^2 + \sin^2\ttheta \, d\varphi^2\right)$, 
$r_h = 2G_{\rm N}m/c^2$ is the Schwarzschild horizon radius 
($G_{\rm N}$ is Newton's constant, $m$ is the black-hole mass, and 
$c$ is the speed of light; for ease of notation, we use natural units 
$c=\hbar=1$ throughout), and $\kappa = 1/(4G_{\rm N}m)$ is the corresponding  surface gravity and
\begin{eqnarray}
\rho_0 \ = \ \sqrt{8G_{\rm{N}}m \ \! (r -  r_h)}\, ,
\end{eqnarray}
represents the proper radial distance from the horizon (in the near-horizon approximation).

The metric in Eq.~(\ref{B.1.kh}) relates the Killing energy $E_{\infty}$
to the locally measured energy $E_{\rm loc}$
of a static observer at proper distance $\rho_0$ from the horizon. Explicitly
\begin{equation}
E_{\rm loc} \ = \  \frac{E_{\infty}}{\kappa\rho_0}\, .
\label{B.3.kl}
\end{equation}
For a photon, whose energy is related to its frequency by $E = h\nu$, this implies
that the locally measured frequency is
\begin{equation}
\nu_{\rm loc}
\ = \ \frac{\nu_{\infty}}{\kappa \rho_0}
= \frac{\nu_{\infty}}{\sqrt{1 - \frac{2G_{\rm N} m}{r}}}\, .
\label{B.4.ji}
\end{equation}
As $r \to 2G_{\rm N} m$, the locally measured frequency diverges, reflecting the
infinite blueshift experienced by a photon approaching the black-hole horizon.
Eq.~(\ref{B.4.ji}) thus encodes the standard gravitational redshift (or
blueshift) of photon frequencies in the stationary Schwarzschild spacetime
\cite{Wald:84}.

Consider a small cavity of proper volume $V_{\rm prop} = A\, \Delta \rho$.  For an infinitesimal slab of the cavity with radial thickness $d\rho$, the
corresponding proper volume is $dV_{\rm prop} = A\, d\rho$. The number of single-particle photon states whose locally measured energy
lies in the interval $[E_{\rm loc},\, E_{\rm loc}+dE_{\rm loc}]$ is then given by
\begin{eqnarray}
&&\mbox{\hspace{-10mm}}g_{\rm loc}(E_{\rm loc})\, dE_{\rm loc}\nonumber \\[2mm]
&&= \ 
2\,\frac{dV}{(2\pi)^3}
\int d^3\pmb{p}\, \delta\!\left(|\pmb{p}|-E_{\rm loc}\right)\, dE_{\rm loc}\, ,
\end{eqnarray}
where the factor 2 is due to two physical photon polarizations.
The momentum integral gives 
\begin{equation}
g_{\rm loc}(E_{\rm loc})
\ = \ 
\frac{dV}{\pi^2}\, E_{\rm loc}^2\, .
\end{equation}
Equivalently, the cumulative number of local photon modes with energy below
$E_{\rm loc}$ per radial slice of thickness $d\rho$ is
\begin{equation}
dN_{\rm loc}(E_{\rm loc})
\ = \ 
\int_0^{E_{\rm loc}} g_{\rm loc}(E')\, dE'
\ = \ 
\frac{dV}{3\pi^2}\, E_{\rm loc}^3\, .
\end{equation}
Consequently, the total number of single-particle modes (SPM) with asymptotic
energy below $E_{\infty}$ contained in a cavity located at the proper
distance $\rho_0$ from the horizon is given by
\begin{eqnarray}
&&\mbox{\hspace{-7mm}}\Omega_{\rm SPM}(E_{\infty},\rho_0,\Delta\rho)\nonumber \\[2mm]
 &&\mbox{\hspace{15mm}}=  \
\int_{\rho_0}^{\rho_0+\Delta\rho} dN_{\rm loc}(\rho)\nonumber \\[2mm]
&&\mbox{\hspace{15mm}}=   \
\frac{A E_{\infty}^3}{3\pi^2\kappa^3}
\int_{\rho_0}^{\rho_0+\Delta\rho} \frac{d\rho}{\rho^3} \nonumber \\[2mm]
&&\mbox{\hspace{15mm}}=\
\frac{A E_{\infty}^3}{6\pi^2\kappa^3}
\left(
\frac{1}{\rho_0^2}
\ - \
\frac{1}{(\rho_0+\Delta\rho)^2}
\right)\!.~~~~~
\label{B.8.cv}
\end{eqnarray}
Near the horizon, when $\rho_0 \ll \Delta \rho$, the integral is clearly dominated by $\rho_0$  since
\begin{eqnarray}
\left(
\frac{1}{\rho_0^2}
\ - \
\frac{1}{(\rho_0+\Delta\rho)^2}
\right) \ &=& \ \frac{1}{\rho_0^2} \left[1 \ + \ \mathcal{O}(\rho_0/\Delta \rho)^2   \right] \nonumber \\[2mm]  &\simeq& \ \frac{1}{\rho_0^2}\, , 
\end{eqnarray}
hence 
\begin{eqnarray}
\Omega_{\rm SPM}(E_{\infty},\rho_0,\Delta\rho) \ = \ \frac{A E_{\infty}^3}{6\pi^2\kappa^3 \rho_0^2}\, .
\label{b.10.ka}
\end{eqnarray}
This area-law scaling reflects the blueshift dominance of the radiation near to horizon.  
In particular,  the majority of states below a given Killing energy $E_{\infty}$ are concentrated in a thin near-horizon layer.

Let us now turn to the many-body state space scaling. To this end, we denote the spectrum of single-particle energies of a free bosonic field (here,
photons) in a fixed background as $\{\varepsilon_i\}_{i=1}^{\infty}$. The cumulative single-particle mode
counting function, discussed so far, is defined as
\begin{equation}
\Omega_{\rm SPM}(E,\rho_0,\Delta\rho)  \ \equiv \ \# \{\, i \mid \varepsilon_i \le E \,\}\, .
\end{equation}
The corresponding multi-particle bosonic Fock space is
\begin{eqnarray}
\mathcal{F}
\  = \ 
\bigoplus_{n=0}^{\infty} \ \!
\mathrm{Sym}\!\left(\mathcal{H}_1^{\otimes n}\right)\, ,
\end{eqnarray}
where $\mathcal{H}_1$ is the single-particle Hilbert space.
%
%

We now define the density of states with exactly $n$ photons and total energy $E$ as 
$\omega_n(E)$.
The total {\em microcanonical} density of states with energy $E$ is then
\begin{eqnarray}
\omega(E) \ = \  \sum_{n=0}^{\infty} \omega_n(E)\, .
\end{eqnarray}
Ensuing {\em cumulative} number of states with {\em total} energy less than or equal to $E$ is
\begin{eqnarray}
\Omega_{\rm{TC}}(E) \ = \  \int_{0}^{E} \omega(E')\, dE'\, .
\end{eqnarray}

A many-photon state in the bosonic Fock space is characterized by a set
of occupation numbers \(\{m_i\}\) of the single-particle modes
\(\{\epsilon_i\}\), satisfying
\begin{eqnarray}
\sum_i m_i \ = \ n\, ,
\qquad
\sum_i m_i\epsilon_i \ = \ E\, ,
\end{eqnarray}
where \(m_i=0,1,2,\ldots\). Taking into account both the
indistinguishability of photons and the possibility of multiple
occupation of a single mode, the density of states with exactly \(n\)
photons satisfies the recursion relation
\begin{eqnarray}
\omega_n(E)
\ = \
\frac{1}{n}
\sum_{k=1}^{n}
\frac{1}{k}
\int_0^E d\epsilon\,
g_1(\epsilon/k)\,
\omega_{n-k}(E-\epsilon)\, .~~~~~
\label{B.16.kl}
\end{eqnarray}
Here, $k$ denotes the number of photons occupying the same
single-particle mode. The term with \(k=1\) corresponds to singly
occupied modes, while the terms with \(k>1\) account for multiple
occupation of a mode, which is the characteristic feature of Bose
statistics. The factor \(g_1(\epsilon/k)\) appears because a mode occupied by
\(k\) identical photons has total energy \(\epsilon\), while the
corresponding one-photon energy is \(\epsilon/k\). The factor \(1/k\)
is the combinatorial weight associated with the \(k\)-fold occupation of
a bosonic mode. The overall factor \(1/n\) arises when extracting the
contribution with exactly \(n\) photons from the generating function of
the bosonic Fock space.

Since, by definition, $\omega_1(E) = g_1(E)$, we see that $g_1(\epsilon)$ describes one-particle density of states with energy $\epsilon$.
To obtain $g_1(\epsilon)$, we use the fact that the cumulative number of one-particle modes with asymptotic energy below $\epsilon = E_\infty$ has already been obtained in Eq.~(\ref{b.10.ka}) by integrating the local density over the proper volume of the cavity. The ensuing one-particle density of states is therefore
\begin{equation}
g_1(\epsilon;\rho_0,\Delta\rho)
\ = \ 
\frac{\partial}{\partial \epsilon}
\Omega_{\rm SPM}(\epsilon;\rho_0,\Delta\rho)\ ,
\label{B14a}
\end{equation}
where the dependence on $\rho_0$ (and $\Delta\rho$) is purely parametric, reflecting the position and geometry of the cavity.

Note also that, since  $\omega_1(E) = g_1(E)$,  we have 
\begin{eqnarray}
\omega_0(E) \ = \  \delta(E)\, .
\label{B.18.df}
\end{eqnarray}
To evaluate the total density of states, it is convenient to introduce
the (one-sided) Laplace transform of $g_1$:
\begin{equation}
\tilde{g}(s) \ = \  \mathcal{L}\{g_1\}(s) \ = \  \int_0^{\infty} e^{-s\epsilon} g_1(\epsilon)\, d\epsilon\, , 
\end{equation}
where $\Re(s)  >   0$.
This allows to rewrite the recursion relation~(\ref{B.16.kl}) in the form
\begin{eqnarray}
\mathcal{L}\{\omega_n\}(s) \ = \ \frac{1}{n} \sum_{k=1}^n \tilde{g}(ks) \ \! \mathcal{L}\{\omega_{n-k}\}(s)\, .
\end{eqnarray} 
This recursion might be solved by defining a generating function (analogue of a fugacity expansion)   
\begin{eqnarray}
\mathcal{W}(x,s) \ = \  \sum_{n=0}^\infty \mathcal{L}\{\omega_n\}(s) \, x^n\, .
\end{eqnarray}
It can be checked that $\mathcal{W}(x,s)$ satisfies the Euler-type 
partial differential equation
\begin{eqnarray}
x \frac{\partial \mathcal{W}}{\partial x}(x,s) \ = \  \mathcal{H}(x,s) \, \mathcal{W}(x,s)\, ,
\label{B.22.jh}
\end{eqnarray}
where
\begin{eqnarray}
\mathcal{H}(x,s) \ = \  \sum_{k=1}^\infty \tilde{g}(ks) \, x^k\, .
\end{eqnarray}
The solution of~(\ref{B.22.jh}) has the form
\begin{eqnarray}
\mathcal{W}(x,s) \ = \ \mathcal{W}_0(s) \exp\left[ \int_0^x \frac{\mathcal{H}(u,s)}{u} \, du \right]\, .
\end{eqnarray}
By employing the fact, cf. Eq.~(\ref{B.18.df}), that $\mathcal{W}(0,s) = 1$, we get that $\mathcal{W}_0(s) = 1$.
So, finally, we obtain 
\begin{eqnarray}
\mathcal{W}(x,s)
\ = \ 
\exp\left[
\sum_{k=1}^{\infty}
\frac{x^k}{k}\tilde{g}(ks)
\right]\, .
\end{eqnarray}
Consequently, after summing over all photon numbers, i.e. setting $x=1$,
the cumulative number of states can be written as
\begin{widetext}
\begin{eqnarray}
\Omega_{\rm TC}(E)
\ = \ 
\frac{1}{2\pi i}
\int_{c_0-i\infty}^{c_0+i\infty}
ds\,
\exp\left[
sE \ + \ 
\sum_{k=1}^{\infty}
\frac{1}{k}\tilde{g}(ks)
\right].
\label{B.22.jkl}
\end{eqnarray}
\end{widetext}
The parameter $c_0$ in the inverse Laplace transform is chosen so that the (Bromwich) integration contour $Re(s) = c_0$ lies to the right of all singularities of $\tilde{g}(s)$.

So, in particular, from Eq.~(\ref{B14a}) we see that
$g_1(\epsilon)=\zeta\,\epsilon^2$, where
$\zeta=A/(2\pi^2\kappa^3\rho_0^2)$.
Hence, $\tilde{g}(s)=2\zeta/s^3$,
and the integration contour may thus be chosen with $c_0>0$.

To evaluate ~ (\ref{B.22.jkl}) in thermodynamic limit, we 
assume that $A$ is large and that  energy density is constant, i.e.
\begin{eqnarray}
\frac{E_\infty}{A\Delta \rho_\infty} \ = \ u_{\infty} \ = \  {\mbox{const}}\, .
\end{eqnarray}
For large $A$ (or equivalently $E_{\infty}$), the integral ~(\ref{B.22.jkl}) is dominated by the saddle point of 
\begin{eqnarray}
\phi(s)
\ &=& \
sE \ + \ 
\sum_{k=1}^{\infty}\frac{1}{k}\tilde{g}(ks)
 \ = \
sE \ + \ 
\frac{2\zeta}{s^3}
\sum_{k=1}^{\infty}\frac{1}{k^4}
\nonumber\\[2mm]
&=&
sE \ + \ \frac{\zeta\pi^4}{45s^3}\, .
\end{eqnarray}
The steepest-descent condition $\phi'(s_{*}) = 0$ gives  $s_*=\pi\left({\zeta}/{15E}\right)^{1/4} $ and so at the saddle
\begin{eqnarray}
\phi(s_*) \ &=& \ E s_* \ + \ \frac{\zeta \pi^4}{15 s_*^3} \ = \ \frac{4}{3} \left( \frac{\zeta \pi^4 E^3}{15} \right)^{1/4} \nonumber \\[2mm] 
&=& \ \frac{4}{3} A \left( \frac{\pi^{2} }{30 \kappa^{3} \rho_{0}^{2}} \right)^{1/4} (u_{\infty} \Delta \rho_{\infty})^{3/4} \ .
\end{eqnarray}
So, in thermodynamic limit we have
\begin{eqnarray}
\Omega_{\rm TC}(E_{\infty},\rho_0,\Delta \rho )\sim  \exp\left[\frac{4}{3} A \left( \frac{\pi^{2} }{30 \kappa^{3} \rho_{0}^{2}} \right)^{1/4} \!\sigma_{\infty}^{3/4}\right] ,~~~~~
\end{eqnarray}
where $\sigma_{\infty}$ is energy per unit transverse area (or surface energy density) of the box at infinity. We have neglected the Gaussian fluctuation factor arising in the saddle-point approximation, since it contributes only a subdominant logarithmic correction to the leading result.


\subsection{Region far from the horizon}

Away from the  horizon, but still within the near-horizon regime, i.e., for $\Delta \rho \ll \rho_0 \ll r_h$, the scaling of $\Omega_{\rm SPM}(E_{\infty},\rho_0,\Delta\rho)$ becomes
\begin{eqnarray}
\Omega_{\rm SPM}(E_{\infty},\rho_0,\Delta\rho) \ &=& \ \frac{A E_{\infty}^3}{6\pi^2\kappa^3}
\left(
\frac{1}{\rho_0^2}
\ - \
\frac{1}{(\rho_0+\Delta\rho)^2}
\right)\nonumber \\[2mm]
&=&\ \frac{V_{\rm prop} E_{\infty}^3}{3\pi^2(\kappa\rho_0)^3} \left[1 \ + \ \mathcal{O}(\Delta \rho/\rho_0)\right]\nonumber\\[2mm]
&\simeq& \ \frac{V_{\rm prop} E_{\infty}^3}{3\pi^2(\kappa\rho_0)^3}\, .
\end{eqnarray}
So, if we move the cavity away from the horizon the blueshift decreases, and $\Omega_{\rm SPM}(E_{\infty},\rho_0,\Delta\rho)$ scales with the proper volume.

In the asymptotically flat region, $r\gg r_h$, the near-horizon 
approximation is no longer applicable and the redshift factor
$\kappa\rho_0$ in~(\ref{B.3.kl}) is replaced by the exact Schwarzschild factor
$\sqrt{-g_{00}}$. In this limit
\begin{eqnarray}
\sqrt{-g_{00}}
\ = \ 
\sqrt{1-\frac{r_h}{r}}
\ \rightarrow \ 1\, ,
\end{eqnarray}
while the proper volume becomes
$
V_{\rm prop}\rightarrow V_{\infty}\equiv V=A\Delta r$.
In this limit, Eq.~(\ref{B.8.cv}) reduces to the standard Minkowski result~\cite{Reif1965}
\begin{eqnarray}
\Omega_{\rm SPM}(E_{\infty},r,\Delta r)
&=&
\frac{V E_{\infty}^3}{3\pi^2}, 
\end{eqnarray}
so that $g_1(\epsilon) = \zeta \epsilon^2$ with $\zeta = V/\pi^2$.

Once we move from single-particle modes to the multi-particle Fock space, the
number of accessible quantum states with total energy less than or equal to $E$
can be, similarly as in the previous subsection, obtained from~(\ref{B.22.jkl}) via  the steepest-descent condition $\phi'(s_{*}) = 0$. In this case 
\begin{eqnarray}
\phi(s_*) \ &=&  \frac{4}{3} \left( \frac{\zeta \pi^4 E^3}{15} \right)^{1/4} \ = \ \frac{4}{3}\left(\frac{\pi^2}{15}\right)^{1/4} u_{\infty}^{3/4}\,V\, ,~~~~~~~
\end{eqnarray}
and thus the leading exponential behaviour is
\begin{eqnarray}
\Omega_{\rm TC}(E_{\infty},\rho_0,\Delta \rho )  \ \sim \ \exp\left[\frac{4}{3}V\left(\frac{\pi^2}{15}\right)^{1/4} u_{\infty}^{3/4} \right]\, . ~~~~
\end{eqnarray}
The latter gives 
the microcanonical entropy (with $\gamma = 1$)
\begin{eqnarray}
S_1 \ = \  \frac{4}{3}\left(\frac{\pi^2}{15}\right)^{1/4} u_{\infty}^{3/4}\,V\, .~~~~~
\end{eqnarray}
When coupled with the condition $E_{\infty} = V u_{\infty}$, we obtain  
\begin{eqnarray}
S_1 \ = \ \frac{4}{3}\left(\frac{\pi^2}{15}\right)^{1/4} V^{1/4} E^{3/4}_{\infty}\, .
\label{B.31.kk}
\end{eqnarray}
This is precisely the result obtained in conventional thermodynamics for EM field. Indeed, there one has (see, e.g.~\cite{L-L:2013})
\begin{eqnarray}
E_{\infty} \ = \ \frac{\pi^2}{15}VT^4, \qquad
S_1 \ = \ \frac{4\pi^2}{45}VT^3\, ,
\end{eqnarray}
and  by eliminating $T$ from these expressions we immediately obtain Eq.~(\ref{B.31.kk}).

%
%
In conclusion, in this appendix we have seen that for a static cavity located outside a black-hole horizon, the phase-space measure associated with field configurations acquires a near-horizon weighting proportional to $\rho^{-2}_0$. This induces a strong
concentration of the modes localized exponentially close to the horizon, so that the effective scaling of the number of accessible states is governed by the  cavity transverse area rather than by the cavity volume. As the cavity is displaced farther from the horizon, the redshift-induced measure concentration diminishes, leading to a crossover from area-dominated  stretched-exponential state-space scaling to ordinary
volume-dominated exponential scaling. An interesting related discussion can be found in Ref.~\cite{Majumdar_2025}.

\section{Computation of Eq.~(\ref{B.110.df}) \label{Appendix-C}}

We consider R\'{e}nyi's entropy~\cite{Renyi} 
\begin{eqnarray}
{{S}}_{\alpha}^{\rm{R}}  \ = \   \frac{1}{1-\alpha} \ \! \log \sum_{i=1}^{W} p_i^{\alpha}\,
,\;\;\;\;\;\;\;\; \alpha > 0\, .\label{II1}
\end{eqnarray}
To determine the extremum, we restrict ourselves for simplicity to the canonical-ensemble analogue, in which the prior information is specified by a fixed expectation value of the energy.
The corresponding MaxEnt distributions for ${{S}}_{\alpha}^{\rm{R}}$  can be obtained by extremization of the
associated inference functional
\begin{eqnarray}
L_{\alpha}^{\rm{R}}[{\mathcal{P}}] \ = \ {{S}}_{\alpha}^{\rm{R}}  \ - \  a
\sum_{i=1}^W p_i  \ - \   b \langle H \rangle\, ,
\end{eqnarray}
where $a$ and  $b$ are Lagrange multipliers, the latter serving as the analogue of the inverse temperature (in natural units).
The expectation value of the energy is defined with respect to usual linear mean as
\begin{equation}
\langle H \rangle \ = \ \sum_{i=1}^W p_i E_i\, .
\label{5.aa}
\end{equation}
In the following, we denote the observed value of the mean energy as $U$.

The MaxEnt distribution arises from the condition $\delta L_{\alpha}^{\rm{R}}[{\mathcal{P}}]/\delta p_k  =  0$. This  yields (after multiplying by $p_k$ and summing over $k$)  an equation for $p_k$, namely
\begin{eqnarray}
\tilde{a} X_k \ - \ b_k  \ - \ 1 \ = \ 0\, ,
\label{6b}
\end{eqnarray}
where,  $X_k =p_k^{\alpha-1} $. The parameters involved read
\begin{eqnarray}
 &&\tilde{a} \ = \ \frac{1}{Z_{\alpha}} \ \equiv \ \frac{1}{\sum_{l=1}^W p_l^{\alpha}}\, , \nonumber \\[2mm]
 &&b_k \ = \ b \ \! \frac{(1-\alpha)[E_k \ - \ U ]}{{\alpha}} \, .
\end{eqnarray}
We note in passing that the solution of Eq.~(\ref{6b}) takes the form
\begin{eqnarray}
p_k \ =  \ p_k(Z_{\alpha}, b,  U; E_k)\, .
\end{eqnarray}
With two constraints at hand we could eliminate $Z_{\alpha}$    and rewrite $b$ in terms of $U$ (or, more physically,  $U$ in terms of $b$). In fact, the solution of~(\ref{6b}) directly yields
the MaxEnt distribution for R\'{e}nyi's entropy 
\begin{eqnarray}
{p}_k  \ = \  Z^{-1}\left[ 1  - \beta^{\rm R} (\alpha-1) \Delta E_k \right]^{{1}/{(\alpha-1)}}\, . \label{2.7}
\end{eqnarray}
Here $\Delta E_k = E_k - U$, and $Z = Z_{\alpha}^{1/(1-\alpha)}$ is the normalization constant
(basically the partition function). The term \mbox{$\beta^{\rm R} = {{b }/{\alpha}}$} is the analogue of the inverse
``temperature'' of the system. Connection of $\beta^{\rm{R}}$ to absolute temperature is discussed in Sec.~\ref{VII.B.kk}.

\section{Energy density of electromagnetic radiation in a cavity in a static spacetime \label{appendix:D}}

In this appendix we prove Eq.~(\ref{124.hk}).
To this end, we consider an electromagnetic field confined in a cavity in the exterior region of a static black-hole spacetime. Let $\xi^\mu=(\partial_t)^\mu = (\partial_0)^\mu$ denote the timelike Killing vector associated with stationarity, normalized such that
\begin{eqnarray}
\xi^\mu \xi_\mu \ \to \  -1
\quad
\text{as}
\quad
r \ \to \  \infty\, .
\end{eqnarray}
The electromagnetic stress-energy tensor is
\begin{eqnarray}
T_{\mu\nu}
\ = \ 
\frac{1}{4\pi}
\left(
F_{\mu\alpha} F_{\nu}{}^{\alpha}
\ - \ 
\frac{1}{4} g_{\mu\nu} F_{\alpha\beta} F^{\alpha\beta}
\right),
\end{eqnarray}
and satisfies local covariant conservation
\begin{eqnarray}
\nabla_\mu T^{\mu\nu} \ = \  0\,  .
\end{eqnarray}
The energy density measured by an observer with four-velocity $v^\mu$ is defined covariantly as
\begin{eqnarray}
u \ = \  T_{\mu\nu} v^\mu v^\nu\, .
\end{eqnarray}
Static observers follow the integral curves of the Killing field, with four-velocity
\begin{eqnarray}
v^\mu
\ = \ 
\frac{\xi^\mu}{\sqrt{-\xi^\alpha \xi_\alpha}}
\ = \ 
\frac{\xi^\mu}{\sqrt{-g_{00}}}\, .
\end{eqnarray}
Because the spacetime is static, the electromagnetic field can be expanded in modes with a definite Killing frequency $\nu$ associated with $\xi^\mu$. For each mode, the locally measured energy is redshifted according to [see Eq.~(\ref{B.4.ji})]
\begin{eqnarray}
\nu_{\rm local} \ = \ \frac{\nu_\infty}{\sqrt{-g_{00}}}\, .
\end{eqnarray}
The energy density is obtained by summing over all modes. Since the phase-space volume element for photons is conserved under parallel transport along Killing trajectories (Liouville's theorem in curved spacetime~\cite{MTW}), the number of modes per unit proper volume scales as $\sim (-g_{00})^{-3/2}$. Each mode carries energy $\nu_{\rm local} \sim (-g_{00})^{-1/2}$. Multiplying these factors, the total local energy density scales as
\begin{eqnarray}
u(r) \ = \ \frac{u_\infty}{(-g_{00}(r))^2}\, .
\end{eqnarray}
Note that this derivation uses  the properties of the electromagnetic field in a static cavity and the redshift of mode energies.  In particular, it does not invoke the thermodynamic equation of state.

\bibliographystyle{unsrt}

\begin{thebibliography}{10}

\bibitem{Huang}
K.~Huang.
\newblock {\em Statistical Mechanics}.
\newblock John Wiley \& Sons, Inc., New Jersey, 1987.

\bibitem{Reif1965}
F.~Reif.
\newblock {\em Fundamentals of Statistical and Thermal Physics}.
\newblock McGraw-Hill, New York, 1965.

\bibitem{Boltzmann}
L.~Boltzmann.
\newblock Weitere Studien {\"u}ber das W{\"a}rmegleichgewicht unter Gasmolek{\"u}len.
\newblock {\em Sitzungsberichte Akademie der Wissenschaften}, 66:275--370, 1872.

\bibitem{Gibbs}
J.W. Gibbs.
\newblock {\em Elementary Principles in Statistical Mechanics}.
\newblock Charles Scribner's Sons, New York, 1902.

\bibitem{Shannon}
C.~Shannon.
\newblock A {Mathematical} {Theory} of {Communication}.
\newblock {\em Bell Syst. Tech. J.}, 27:379--423,623--656, 1948.

\bibitem{Jaynes}
E.T. Jaynes.
\newblock {\em Papers on Probability, Statistics, and Statistical Physics}.
\newblock D. Reidel Publishing Co, Dordrecht, Holland, 1983.

\bibitem{JPPT2018}
H.J. Jensen, R.H. Pazuki, G.~Pruessner, and P.~Tempesta.
\newblock Statistical mechanics of exploding phase spaces: ontic open systems.
\newblock {\em J. Phys. A: Math. Theor.}, 51:375002, 2018.

\bibitem{HanelTHurner2011}
R.~Hanel and S.~Thurner.
\newblock When do generalized entropies apply? {How} phase space volume determines entropy.
\newblock {\em epl}, 96:50003, 2011.

\bibitem{Frautschi1971}
S.~Frautschi.
\newblock Statistical bootstrap model of hadrons.
\newblock {\em Physical Review D}, 3:2821--2834, 1971.

\bibitem{GreenSchwarzWitten1987}
Michael~B. Green, John~H. Schwarz, and Edward Witten.
\newblock {\em Superstring Theory, Volume 1: Introduction}.
\newblock Cambridge University Press, Cambridge, UK, 1987.

\bibitem{Polchinski1998}
Joseph Polchinski.
\newblock {\em String Theory, Volume 1: An Introduction to the Bosonic String}.
\newblock Cambridge University Press, Cambridge, UK, 1998.

\bibitem{BandtPompe2002}
C.~Bandt and B.~Pompe.
\newblock Permutation entropy: A natural complexity measure for time series.
\newblock {\em Phys. Rev. Lett.}, 88:174102, 2002.

\bibitem{Hanel}
S.~Thurner, R.~Hanel, and P.~Klimek.
\newblock {\em Introduction to the Theory of Complex Systems}.
\newblock Oxford University Press, Oxford, 2019.

\bibitem{Tsac}
C.~Tsallis.
\newblock {\em Introduction to Nonextensive Statistical Mechanics; Approaching a Complex World}.
\newblock Springer, New York, 2009.

\bibitem{Tempesta}
P.~Tempesta.
\newblock Multivariate group entropies, super-exponentially growing complex systems, and functional equations.
\newblock {\em Chaos}, 30:123119, 2020.

\bibitem{T-J}
C.~Tsallis and H.J. Jensen.
\newblock Extensive composable entropy for the analysis of cosmological data.
\newblock {\em Phys. Lett. B}, 861:139238, 2025.

\bibitem{Tsallis:2012js}
C.~Tsallis and J.L. Cirto.
\newblock Black hole thermodynamical entropy.
\newblock {\em Eur. Phys. J. C}, 73:2487, 2013.

\bibitem{Tsab}
C.~Tsallis.
\newblock Black hole entropy: A closer look.
\newblock {\em Entropy}, 22:17, 2020.

\bibitem{Tsallis_epl_2026}
C.~Tsallis.
\newblock Area law: Canonical and microcanonical thermostatistics grounded on nonadditive entropies.
\newblock {\em EPL}, 153:61001, 2026.

\bibitem{BeckTsallis2026}
C.~Beck and C.~Tsallis.
\newblock Turbulence: An entropic approach.
\newblock {\em arXiv:2606.03139}, 2026.

\bibitem{PT2011PRE}
P.~Tempesta.
\newblock Group entropies, correlation laws, and zeta functions.
\newblock {\em Phys. Rev. E}, 84:021121, 2011.

\bibitem{T2016AOP}
P.~Tempesta.
\newblock Beyond the {Shannon}--{Khinchin} formulation: The composability axiom and the universal-group entropy.
\newblock {\em Ann. Phys.}, 365:180--197, 2016.

\bibitem{Tempesta2016}
P.~Tempesta.
\newblock Formal groups and {Z}-entropies.
\newblock {\em Proc. R. Soc. A}, 472:20160143, 2016.

\bibitem{SJ1}
J.E. Shore and R.W. Johnson.
\newblock Axiomatic derivation of the principle of maximum entropy and the principle of minimum cross-entropy.
\newblock {\em IEEE Trans. Inf. Theor.}, 26:26--37, 1980.

\bibitem{SJ2}
J.E. Shore and R.W. Johnson.
\newblock Properties of cross-entropy minimization.
\newblock {\em IEEE Trans. Inf. Theor.}, 27:472--482, 1981.

\bibitem{J-K:19}
P.~Jizba and J.~Korbel.
\newblock Maximum entropy principle in statistical inference: Case for non-Shannonian entropies.
\newblock {\em Phys. Rev. Lett.},
122:120601, 2019.

\bibitem{J-Temp}
H.J. Jensen and P.~Tempesta.
\newblock Group {Entropies}: {From} {Phase} {Space} {Geometry} to {Entropy} {Functionals} via {Group} {Theory}.
\newblock {\em Entropy}, 20:804, 2018.

\bibitem{Caratheodory}
C.~Carath\'{e}odory.
\newblock Untersuchungen {\"u}ber die Grundlagen der Thermodynamik.
\newblock {\em Math. Ann.}, 67:355--386, 1909.

\bibitem{Barrow}
J.~D. Barrow.
\newblock The area of a rough black hole.
\newblock {\em Phys.~Lett. B}, 808:135643, 2020.

\bibitem{JL}
P.~Jizba and G.~Lambiase.
\newblock Constraints on {Tsallis} {Cosmology} from {Big} {Bang} {Nucleosynthesis} and the {Relic} {Abundance} of {Cold} {Dark} {Matter} {Particles}.
\newblock {\em Entropy}, 25:1495, 2023.

\bibitem{JLLM:24}
P.~Jizba, G.~Lambiase, G.G. Luciano, and L.~Mastrototaro.
\newblock Imprints of {B}arrow-{T}sallis cosmology in primordial gravitational waves.
\newblock {\em Eur. Phys. J. C}, 84:1076, 2024.

\bibitem{Majumdar_2025}
E.~Aurell and S.N. Majumdar.
\newblock Operators with spectral density increasing as {$\Omega(E)\sim e^{\rm Const. {E}^2}$}.
\newblock {\em Phys. Rev. Res.}, 7:043165, 2025.

\bibitem{Kchinchin}
A.I. Khinchin.
\newblock {\em Mathematical Foundations of Statistical Mechanics}.
\newblock Dover Publications Inc, New York, 1949.

\bibitem{Johal}
R.S. Johal.
\newblock Composable entropy and deviation from macroscopic equilibrium.
\newblock {\em Phys. Lett. A}, 332:345--349, 2004.

\bibitem{Scarfone}
A.M. Scarfone and T.~Wada.
\newblock Thermodynamic equilibrium and its stability for microcanonical systems described by the {Sharma}-{Taneja}-{Mittal} entropy.
\newblock {\em Phys. Rev. E}, 72:026123, 2005.

\bibitem{Hotta}
M.~Hotta and I.~Joichi.
\newblock Composability and generalized entropy.
\newblock {\em Phys. Lett. A.}, 262:302--309, 1999.

\bibitem{Haze}
M~Hazewinkel.
\newblock {\em Formal Groups and Applications.}
\newblock Academic Press, New York, 1978.

\bibitem{Ding:2008}
L.~Ding, N.~Bray-Ali, R.~Yu, and S.~Haas.
\newblock Subarea law of entanglement in nodal fermionic systems.
\newblock {\em Phys. Rev. Lett.}, 100:215701, 2008.

\bibitem{Li:2006}
W.~Li, L.~Ding, R.~Yu, T.~Roscilde, and S.~Haas.
\newblock Scaling behavior of entanglement in two- and three-dimensional free-fermion systems.
\newblock {\em Phys. Rev. B}, 74:073103, 2006.

\bibitem{KLich:2014}
I.~Klich, S.H. Lee, and K.~Iida.
\newblock Glassiness and exotic entropy scaling induced by quantum fluctuations in a disorder-free frustrated magnet.
\newblock {\em Nature Commun.}, 5:3497, 2014.

\bibitem{Andrews:2009}
S.~Andrews, H.~De~Sterck, S.~Inglis, and R.G. Melko.
\newblock Monte {Carlo} study of degenerate ground states and residual entropy in a frustrated honeycomb lattice {Ising} model.
\newblock {\em Phys. Rev. E}, 79:041127, 2009.

\bibitem{Fisch:2006}
R.~Fisch.
\newblock Finite-{Size} {Scaling} in the {Energy}-{Entropy} {Plane} for the 2{D} $\pm$ Ising {S}pin {G}lass.
\newblock {\em J. Stat. Phys}, 125:777--792, 2006.

\bibitem{Bekenstein}
J.D. Bekenstein.
\newblock Black {Hole} and {Entropy}.
\newblock {\em Phys. Rev. D}, 7:2333--2346, 1973.

\bibitem{Hawking}
S.~W. Hawking.
\newblock Particle creation by black holes.
\newblock {\em Commun. Math. Phys.}, 43:199--220, 1975.

\bibitem{Hawking2}
G.W. Gibbons and S.~W. Hawking.
\newblock Cosmological event horizons, thermodynamics, and particle creation.
\newblock {\em Phys. Rev. D}, 15:2738--2751, 1977.

\bibitem{FJL:26}
M.~Figliolia, P.~Jizba, and G.~Lambiase.
\newblock Thermodynamic gravity with non-extensive horizon entropy and topological calibration.
\newblock \em Class. Quantum Grav. 43:165014, 2026.

\bibitem{Estrada:2020}
M.~Estrada and R.~Aros.
\newblock Thermodynamic extended phase space and {P}-{V} criticality of black holes at {P}ure {L}ovelock gravity.
\newblock {\em Eur. Phys. J. C}, 80:395, 2020.

\bibitem{Kubiznak:2023}
D.~Kubiz\v{n}\'{a}k and M.~Li\v{s}ka.
\newblock Can {B}ekenstein's area law prevail in modified theories of gravity?
\newblock {\em Phys. Rev. D}, 108:L121501, 2023.

\bibitem{Renyi}
A.~R\'{e}nyi.
\newblock {\em Selected Papers of Alfr\'{e}d R\'{e}nyi, Vol. 2}
\newblock Akad\'{e}miai Kiad\'{o}, Budapest, 1976.

\bibitem{Uffink:95}
J.~Uffink.
\newblock Can the maximum entropy principle be explained as a consistency requirement?
\newblock {\em Stud. Hist. Phil. Mod. Phys.}, 26:223--261, 1995.

\bibitem{EncisoTempesta2017}
A.~Enciso and P.~Tempesta.
\newblock Uniqueness and characterization theorems for generalized entropies.
\newblock {\em J. Stat. Mech.}, 2017:123101, 2017.

\bibitem{Jaynes1957}
E.~T. Jaynes.
\newblock Information theory and statistical mechanics.
\newblock {\em Phys. Rev.}, 106:620--630, 1957.

\bibitem{Rufo:14}
A.~Campa, T.~Dauxois, D.~Fanelli, and S.~Ruffo.
\newblock {\em Physics of Long-Range Interacting Systems}.
\newblock Oxford University Press, Oxford, 2014.

\bibitem{LB}
D.~Lynden-Bell and R.~Wood.
\newblock The gravo-thermal catastrophe in isothermal spheres and the onset of red-giant structure for stellar systems.
\newblock {\em Monthly Notices of the Royal Astronomical Society}, 138:495, 1968.

\bibitem{Thirring}
W.~Thirring.
\newblock Systems with negative specific heat.
\newblock {\em Zeitschrift f\"{u}r Physik}, 235:339, 1970.

\bibitem{L-L:2013}
L.~D. Landau and E.~M. Lifshitz.
\newblock {\em Statistical Physics: Volume 5}.
\newblock Elsevier, London, 2013.

\bibitem{Abe}
S.~Abe, S.~Mart\'inez, F.~Pennini, and A.~Plastino.
\newblock Nonextensive thermodynamic relations.
\newblock {\em Phys. Lett. A}, 281:126--130, 2001.

\bibitem{Biro}
T.~S. Bir{\'o} and P.~V{\'a}n.
\newblock Zeroth law compatibility of nonadditive thermodynamics.
\newblock {\em Phys. Rev. E}, 83:061147, 2011.

\bibitem{TJ2020SR}
P.~Tempesta and H.~J. Jensen.
\newblock Universality {C}lasses and {I}nformation-{T}heoretic {M}easures of {C}omplexity via {G}roup {E}ntropies.
\newblock {\em Sci. Rep.}, 10:5952, 2020.

\bibitem{CaratheodoryII}
H.~A. Buchdahl.
\newblock {O}n the {U}nrestricted {T}heorem of {C}arath{\'e}odory and {I}ts {A}pplication in the {T}reatment of the {S}econd {L}aw of {T}hermodynamics.
\newblock {\em Am. J. Phys.}, 17:212--218, 1949.

\bibitem{Note1}
The thermodynamic volume, denoted by $V_{\protect \rm th}$, is the thermodynamic variable conjugate to the pressure (equivalently, the cosmological constant) in extended black-hole thermodynamics~\cite {Kastor:2009wy,Kubiznak:2016qmn}. It is therefore a thermodynamic response function characterizing how the black-hole enthalpy changes under variations of the cosmological constant, rather than a measure of a spatial region. By contrast, a ``geometric interior volume'' of a black hole is not uniquely defined in general relativity, as it depends on the choice of spacelike foliation and the prescription used to define the interior~\cite {Christodoulou:2014yia}. Nevertheless, for static, spherically symmetric black holes, $V_{\protect \rm th}$ coincides numerically with the Euclidean volume $(4/3)\pi r_{\protect \rm h}^3$ enclosed by the horizon. This distinction is important for our analysis, since, as in conventional statistical thermodynamics, the scaling of the number of states should be formulated in terms of
  a well-defined thermodynamic state variable rather than a foliation-dependent geometric construction.

\bibitem{Capozziello2025}
S.~Capozziello and M.~Shokri.
\newblock Barrow entropies in black hole thermodynamics.
\newblock {\em European Physical Journal C}, 85:200, 2025.

\bibitem{Volovik2026}
G.E. Volovik.
\newblock Hawking radiation: black hole vs de {S}itter.
\newblock {\em arXiv2510.24502v7}, 2026.

\bibitem{Sadeghi2026}
J.~Sadeghi, B.~Pourhassan, and F.~Khosravani.
\newblock Beyond entropy extension and exploring the thermodynamics of {A}d{S} black holes.
\newblock {\em Nucl. Phys. B}, 1025:117382, 2026.

\bibitem{Ahmed2025}
F.~Ahmed, A.~Al-Badawi, and T.~Sakalli.
\newblock Exploring geodesics, quantum fields and thermodynamics of {Schwarzschild-AdS} black hole with a global monopole in non-commutative geometry.
\newblock {\em Nucl. Phys. B}, 1017:116951, 2025.

\bibitem{Nakarachinda2025}
R.~Nakarachinda, C.~Promsiri, and L.~Tannukij.
\newblock Thermodynamics of black holes with {R{\'e}nyi} entropy from classical gravity.
\newblock {\em Nucl. Phys. B}, 1011:116796, 2025.

\bibitem{Bhattacharjee2025}
A.~Bhattacharjee and P.~Phukon.
\newblock Thermodynamics of flat 4{D} {Einstein}-{Gauss}-{Bonnet} black hole with {R{\'e}nyi} entropy: {A}n {R}{P}{S}{T}-like formalism.
\newblock {\em Nucl. Phys. B}, 1019:117103, 2025.

\bibitem{JKZ}
P.~Jizba, J.~Korbel, and V.~Zatloukal.
\newblock Tsallis thermostatics as a statistical physics of random chains.
\newblock {\em Phys. Rev. E}, 95:022103, 2017.

\bibitem{Beck2009}
C.~Beck.
\newblock Generalised information
and entropy measures in physics. \em Contemp. Physics,
50:495, 2009.

\bibitem{Wald:84}
R.M. Wald.
\newblock {\em General Relativity}.
\newblock University of Chicago Press, London, 1984.

\bibitem{tolman}
R.~C. Tolman and P.~Ehrenfest.
\newblock Temperature equilibrium in a static gravitational field.
\newblock {\em Phys. Rev.}, 36:1791--1798, 1930.

\bibitem{tHooft1}
G.~'t~Hooft.
\newblock On the quantum structure of a black hole.
\newblock {\em Nucl. Phys. B}, 256:727--745, 1985.

\bibitem{tHooft2}
G.~'t~Hooft.
\newblock The scattering matrix approach for the quantum black hole: an overview.
\newblock {\em Int. J. Mod. Phys. A}, 11:4623--4688, 1996.

\bibitem{Jizba-Dun:16}
P.~Jizba, Y.~Ma, A.~Hayes, and J.A. Dunningham.
\newblock One-parameter class of uncertainty relations based on entropy power.
\newblock {\em Phys. Rev. E.}, 93:060104(R), 2016.

\bibitem{McQuarrie}
D.A. McQuarrie.
\newblock {\em Statistical Mechanics}.
\newblock University Science Books, Sausalito, California, 2000.

\bibitem{Padmanabhan:90}
T.~Padmanabhan.
\newblock Statistical mechanics of gravitating systems.
\newblock {\em Phys. Rep.}, 188:285--362, 1990.

\bibitem{PhysRevD.111.024013}
A.~Belfiglio, S.~M. Chandran, O.~Luongo, and S.~Mancini.
\newblock Horizon entanglement area law from regular black hole thermodynamics.
\newblock {\em Phys. Rev. D}, 111:024013, 2025.

\bibitem{PhysRevD.104.084030}
S.~Nojiri, S.~D. Odintsov, and V.~Faraoni.
\newblock Area-law versus R\'enyi and Tsallis black hole entropies.
\newblock {\em Phys. Rev. D}, 104:084030, 2021.

\bibitem{PhysRevD.102.064014}
Ch. Promsiri, E.~Hirunsirisawat, and W.~Liewrian.
\newblock Thermodynamics and {V}an der {W}aals phase transition of charged black holes in flat spacetime via {R}\'enyi statistics.
\newblock {\em Phys. Rev. D}, 102:064014, 2020.

\bibitem{SUCU2026170494}
E.~Sucu, S.~Dengiz, and \.{I}. Sakall\i.
\newblock Quantum-corrected thermodynamics of conformal {W}eyl gravity black holes: {GUP} effects and phase transitions.
\newblock {\em Annals of Physics}, 490:170494, 2026.

\bibitem{MTW}
C.W. Misner, K.S. Thorne, and J.A. Wheeler.
\newblock {\em Gravitation}.
\newblock W.H.~Freeman, New York, 1973.

\bibitem{Kastor:2009wy}
David Kastor, Sourya Ray, and Jennie Traschen.
\newblock {Enthalpy} and the {Mechanics} of {AdS} {Black} {Holes}.
\newblock {\em Classical and Quantum Gravity}, 26:195011, 2009.

\bibitem{Kubiznak:2016qmn}
D.~Kubiz\v{n}\'ak, R.~B. Mann, and M.~Teo.
\newblock Black {Hole} {Chemistry}: {Thermodynamics} with {Lambda}.
\newblock {\em Classical and Quantum Gravity}, 34(6):063001, 2017.

\bibitem{Christodoulou:2014yia}
M.~Christodoulou and C.~Rovelli.
\newblock How {Big} is a {Black} {Hole}?
\newblock {\em Phys. Rev. D}, 91(6):064046, 2015.

\end{thebibliography}

\end{document}